\documentclass[aps,twocolumn,prl,10pt,amsmath,amssymb,nofootinbib,showpacs,superscriptaddress,floatfix]{revtex4-1}

\newcommand{\rs}{\rm\scriptscriptstyle}
\usepackage{graphicx}
\usepackage{color}
\usepackage[usenames,dvipsnames]{xcolor}
\usepackage[colorlinks=true,linkcolor=Red,citecolor=Green,linktoc=page]{hyperref}
\usepackage{multirow}
\usepackage{float}
\usepackage{flushend}
\usepackage{balance}
\usepackage[varg]{txfonts}
\usepackage{ulem}
\usepackage{fancyhdr}

\begin{document}

\title{Scaling Universality at the Dynamic Vortex Mott Transition}
\author{Martijn\,Lankhorst}
\affiliation{MESA+ Institute for Nanotechnology, University of Twente,
	7500 AE Enschede, The Netherlands}
\author{Nicola\,Poccia}
\affiliation{Department of Physics, Harvard University, Cambridge, MA 02138, USA}
\author{Martin\,P.\,Stehno}
\affiliation{MESA+ Institute for Nanotechnology, University of Twente,
	7500 AE Enschede, The Netherlands}
\author{Alexey\,Galda}
\affiliation{Materials Science Division, Argonne National Laboratory, 9700 S. Cass Ave, Argonne, IL 60439, USA}
\affiliation{The James Franck Institute and Department of Physics, The University of Chicago, Chicago, IL 60637, USA}
\author{Himadri Barman}
\affiliation{Department of Theoretical Physics, Tata Institute of Fundamental
	Research, Homi Bhabha Road, Navy Nagar, Mumbai 400005, India}
\author{Francesco\,Coneri}
\affiliation{MESA+ Institute for Nanotechnology, University of Twente,
	7500 AE Enschede, The Netherlands}
\author{Hans\,Hilgenkamp}
\affiliation{MESA+ Institute for Nanotechnology, University of Twente,
	7500 AE Enschede, The Netherlands}
\author{Alexander\,Brinkman}
\affiliation{MESA+ Institute for Nanotechnology, University of Twente,
	7500 AE Enschede, The Netherlands}
\author{Alexander\,A.\,Golubov}	
\affiliation{MESA+ Institute for Nanotechnology, University of Twente,
	7500 AE Enschede, The Netherlands}
\affiliation{Moscow Institute of Physics and Technology, Institutskii per. 9, Dolgoprudny, 141700, Moscow District, Russia}	
\author{Vikram\,Tripathi}
\affiliation{Materials Science Division, Argonne National Laboratory, 9700 S. Cass Ave, Argonne, IL 60439, USA}
\affiliation{Department of Theoretical Physics, Tata Institute of Fundamental
	Research, Homi Bhabha Road, Navy Nagar, Mumbai 400005, India}
\author{Tatyana\,I.\,Baturina}
\affiliation{Materials Science Division, Argonne National Laboratory, 9700 S. Cass Ave, Argonne, IL 60439, USA}
\affiliation{A.\,V.\,Rzhanov Institute of Semiconductor Physics SB RAS, 13 Lavrentjev Avenue, Novosibirsk 630090, Russia}
\affiliation{Novosibirsk State University, Pirogova str. 2, Novosibirsk 630090, Russia}
\affiliation{Departamento de F\'isica de la Materia Condensada, Instituto de Ciencia de Materiales Nicol\'as Cabrera and Condensed Matter Physics Center (IFIMAC), Universidad Aut\'onoma de Madrid, 28049 Madrid, Spain}
\author{Valerii\,M.\,Vinokur}
\affiliation{Materials Science Division, Argonne National Laboratory, 9700 S. Cass Ave, Argonne, IL 60439, USA}
\begin{abstract}
The dynamic Mott insulator-to-metal transition (DMT) is key to 
many intriguing phenomena in condensed matter physics yet it 
remains nearly unexplored. 
The cleanest way to observe DMT 
without the interference from disorder and other effects inherent to electronic and atomic systems, is to employ the vortex Mott states formed by
superconducting vortices in a regular array of 
pinning sites.
The applied electric current delocalizes vortices and drives the dynamic vortex Mott transition. 
Here we report the critical behavior of the
vortex system as it crosses the DMT line, driven by either current or temperature. We find  
universal scaling with respect to both, 
expressed by the same scaling function and characterized by a single critical exponent coinciding with the exponent for the thermodynamic Mott transition. 
We develop a theory for the DMT based on the parity reflection-time reversal (${\cal PT}$) symmetry breaking formalism and find that the nonequilibrium-induced Mott transition has the same critical behavior as thermal Mott transition. Our findings demonstrate
the existence of physical systems in which the effect of nonequilibrium drive is to generate effective temperature and hence the transition belonging in the thermal universality class. We establish ${\cal PT}$ symmetry-breaking as a universal mechanism for out-of-equilibrium phase transitions.
\end{abstract}

\maketitle

\section*{Introduction}
A Mott insulator\,\cite{PM:1937,Mott:1949,Mott:1990} is a material that should be a conductor according to the standard band theory of electrical conductivity, but acts as an insulator nonetheless. The Mott insulating state arises because of the concurrent action of electron-electron strong correlations and periodic atomic potential has been always viewed as an exemplary manifestation of many-body quantum physics\,\cite{Sachdev:book}. However, 
 the correspondence between the quantum mechanics of a ${\cal D}$-dimensional system and the classical statistical mechanics of a ${\cal D}+1$-dimensional system\,\cite{Polyakov:1987}, lead to a conjecture that a vortex Mott insulator forms in a type II superconductor if the density of superconducting vortices matches the density of the pinning sites\,\cite{Nelson}. The vortex Mott insulator, albeit purely classical formation, harbors all essential features of its ${2\cal D}$ quantum electronic parent: it is incompressible and vortices remain localized at low temperatures. The existence of the vortex Mott insulator was conclusively evidenced in\,\cite{Zeldov:2009} by measurements of the compressibility of the vortex system localized by the periodic surface holes. Importantly, the observed Mott insulator is actually a ${2\cal D}$ \textit{classical} formation, which therefore can be viewed as the commensurate vortex state that was extensively discussed in the vortex community in terms of the enhanced pinning at matching magnetic fields, see, for example,\,\cite{Moshchalkov:1998} and references therein. The implications of the existence of the vortex Mott state are far reaching and two-fold. First, it teaches us that Mott physics embraces more than believed before and includes classical systems. Second, it offers an opportunity of studying quantum many-body strongly correlated physics by experiments on more easily accessible classical systems. Recent numerical simulations of the $2{\cal D}$ system of Coulomb-interacting classical particles\,\cite{Louk:2017} that demonstrated critical scaling at the dynamic Mott transition are in a perfect concert with this conclusion.

Indeed, Mott insulating state can be destroyed not only by varying temperature or pressure, but
also by applied driving field delocalizing particles\,\cite{Imada:1998,Balents:2005,Sachdev:book,Quantum:2012,Lee:2006}.    The observation of the current-driven vortex Mott insulator-to-metal  transition in a proximity array\,\cite{Science2015}, where the vortex Mott insulator state forms\,\cite{Nelson,Nelson:1998}, was an enabling discovery in experimental Mott physics. It provided the first tangible
example of a dynamic Mott transition having settled the vortex quantum mechanical mapping on a firm experimental basis. 
That the revealed nonequilibrium critical behavior with respect to the nonequilibrium drive is the same as that of conventional thermal Mott transition with respect to temperature, raises a largely open class of questions. 
Among these is a central issue in condensed matter physics: the generalization of a thermodynamic phase transition to nonequilibrium conditions.  There have been tantalizing reports that in systems where tuning parameters like temperature, pressure, or magnetic field alter the symmetry, the nonequilibrium drive generates the effective temperature and the corresponding transition appears in the conventional thermal universality class\,\cite{Millis:2006,Chtchelk:2009}.
The finding of\,\cite{Science2015} paves the way for further generalizing this conclusion to a wider nontrivial class of phase transitions, which, like the Mott transition, are not accompanied by a change of symmetry, and calls for intensifying experimental study of the DMT and the interchangeability of external drive and temperature when crossing the transition line.

The hallmark of an electronic Mott insulator-to-metal transition derived from the Hubbard model\cite{Hubbard:1963},
which encompasses
the essential physics of the Mott insulator and Mott transition,
is a change in the electronic density of states (DOS) from gapped (insulator) to peaked  (metal) shape\cite{Imada:1998,Rosenberg,Kotliar:2000} near the Fermi level.
Experimentally, the Mott transition can be detected by 
measuring the tunneling differential \textit{conductance} as a function of the particle density
and observing the change from a sharp dip, which reflects depletion of the electronic states at the Fermi level, to a peak, which signals that a Mott metal has formed. 
We build on the fact that a vortex system 
trapped in a regular array of pinning sites is a much purer realization of the Hubbard model than
standard electronic Mott materials.
In the dual vortex system, the quantum particles -- vortex correspondence maps the tunneling differential conductance of particles onto the thermally activated differential resistance, $dV/dI$. Hence dip-to-peak reversal of the latter measured as a function of the magnetic field heralds the vortex Mott transition\,\cite{Science2015}. 
%%%%%%%%%%%%%%%%%%%%%%%%%%%%%%%%%%%%%%%%%%%%%%%%%%%%%%%%%%%%%%%%%%%%%%%%%%%%%%%%%%%%
%%%%%%%%%%%%%%%%%%%%%%%%%%%%%%%%%%%%%%%%%%%%%%%%%%%%%%%%%%%%%%%%%%%%%%%%%%%%%%%%%%%% 
\begin{figure}[b!]
	\begin{center}
		\begin{center}
			\includegraphics[width=0.7\linewidth]{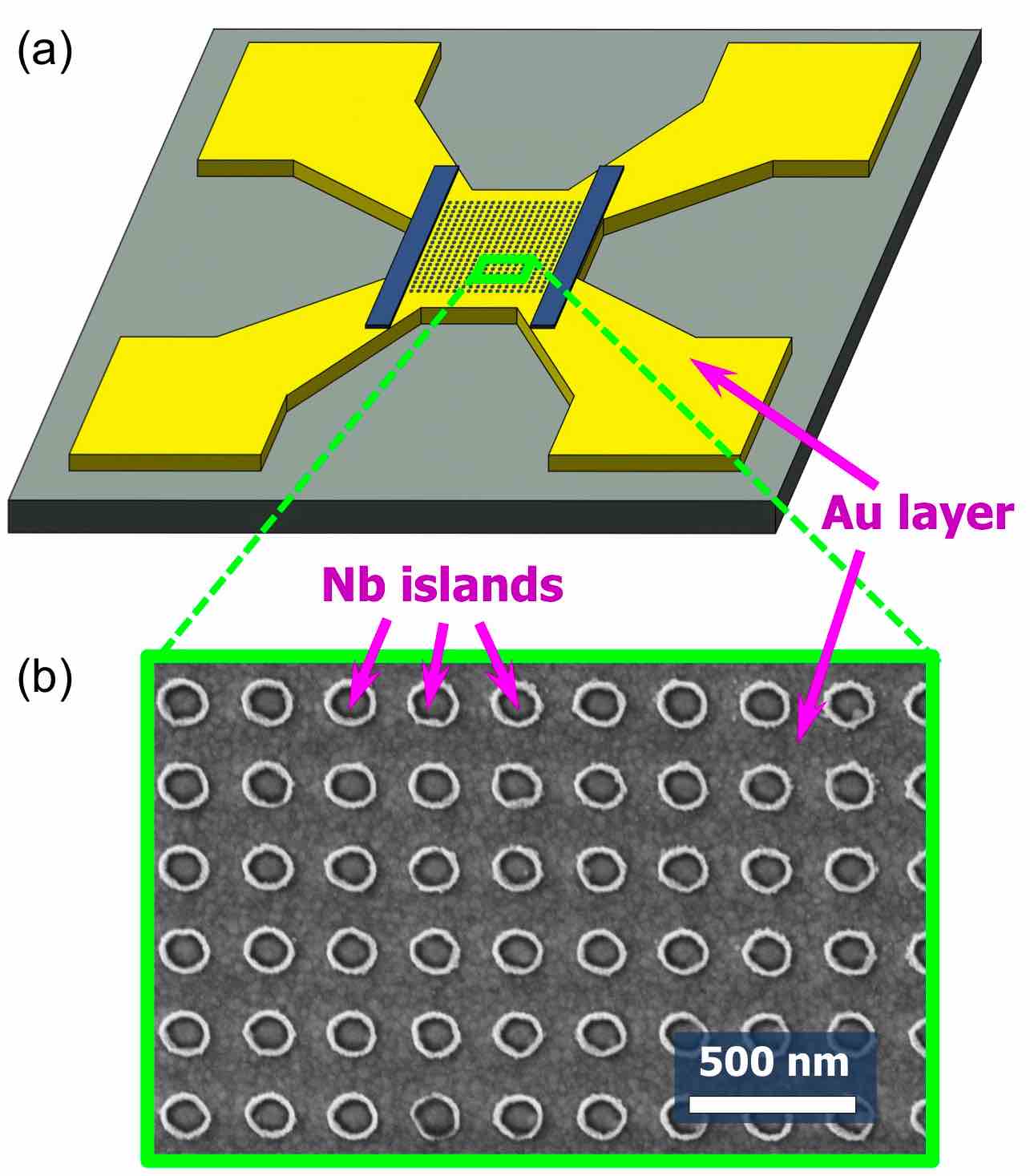}
		\end{center} 
		\caption{Experimental realization of charge-vortex duality for Mott insulator.
			(a) A sketch of the device. The device consists of a square array of 270 $\times$ 270 superconducting Nb islands on a conducting Au layer. On both sides of the array, a Nb bar is placed to ensure the current passes through the array homogeneously. 
			The potential difference between the bars is measured as a function of the external current and an external magnetic field perpendicular to the plane of the array. 
			(b) Scanning electron microscopy image of the sample. 		
		}
		\label{Fig1}
	\end{center}
\end{figure}
%%%%%%%%%%%%%%%%%%%%%%%%%%%%%%%%%%%%%%%%%%%%%%%%%%%%%%%%%%%%%%%%%%%%%%%%%%%%%%%%%%%%
%%%%%%%%%%%%%%%%%%%%%%%%%%%%%%%%%%%%%%%%%%%%%%%%%%%%%%%%%%%%%%%%%%%%%%%%%%%%%%%%%%%% 

\section*{Results}

We focus on the vortex Mott insulator that forms as the vortex matches the density of the regular potential minima, i.e. at the applied magnetic field corresponding to a single flux quantum $\Phi_0=\pi\hbar/e$ per pinning site. 
We create an egg-crate periodic pinning potential 
patterning a square array of $270\times 270$ Nb islands with the lattice constant $a=250$\,nm on a 40 nm-thick base layer of Au on Si/SiO$_2$ as shown in Fig.\,\ref{Fig1}. The islands are $45$\,nm in height and $142 \pm 5$\,nm in diameter. Additionally we placed Nb bars on either sided of the array structure to ensure uniform current injection.
The superconducting transition temperature of the array, $T_c= 2.8$\,K, is determined as the
midpoint of the transition in temperature-resistance curve. 

For our square array, the magnetic field at which the number of vortices matches the traps is $B_0=\Phi_0/a^2$. 
It is convenient to introduce the vortex filling fraction $f=B/B_0$,
so that $f=1$ corresponds to one vortex per lattice cell.
We measure current-voltage characteristics with small steps in magnetic field and temperature and obtain $dV(f)/dI$ curves by numerical differentiation.
From these data, the phase boundary was determined by tracking the position of the dip-to-peak reversal as a function of current ($I$) and temperature ($T$). The details of measurement technique are given in Supplementary Materials (SM).
Figure\,\ref{Fig2}(a) presents the phase diagram of the Mott states in the $T$-$I$ coordinates summarizing the experimental results of our work. 
Representative sets of $dV/dI$ curves are shown in Fig.\,\ref{Fig2}(b,c).
These data were taken using a standard lock-in technique near the transition with very small steps of 5\,$\mu$T in magnetic field, 0.5\,$\mu$A in current (Fig.\,\ref{Fig2}(b)), and 5\,mK in temperature (Fig.\,\ref{Fig2}(c)).
The isothermal plots of panel (b) display the expected dip-to-peak reversal upon increasing the current.
The separatrix current $I_0$ divides between  
the insulating $I<I_0$ and metallic $I>I_0$ phases. 
Note the asymmetry in the $dV/dI$ behaviors at $f<1$ and at $f>1$.
The loci of $I_0(T)$ yield phase transition lines in Fig.\,\ref{Fig2}(a) for $f<1$ and $f>1$.
Fixing the current $I\lesssim I_0$ and then varying temperature 
yield the similar dip-to-peak reversal behavior, see\,Fig.\,\ref{Fig2}(c).
Subtracting the separatrices from the $dV/dI$ data, yields the fan-like set of curves displayed in Fig.\,\ref{Fig2}(d,e), indicating a transition from insulating (bent down towards $f=1$) to metallic (bent up) behaviours.

%%%%%%%%%%%%%%%%%%%%%%%%%%%%%%%%%%%%%%%%%%%%%%%%%%%%%%%%%%%%%%%%%%%%%%%%%%%%%%%%%%%%
%%%%%%%%%%%%%%%%%%%%%%%%%%%%%%%%%%%%%%%%%%%%%%%%%%%%%%%%%%%%%%%%%%%%%%%%%%%%%%%%%%%% 
\begin{figure*}[t!]
	\begin{center}
		\includegraphics[width=0.9\linewidth]{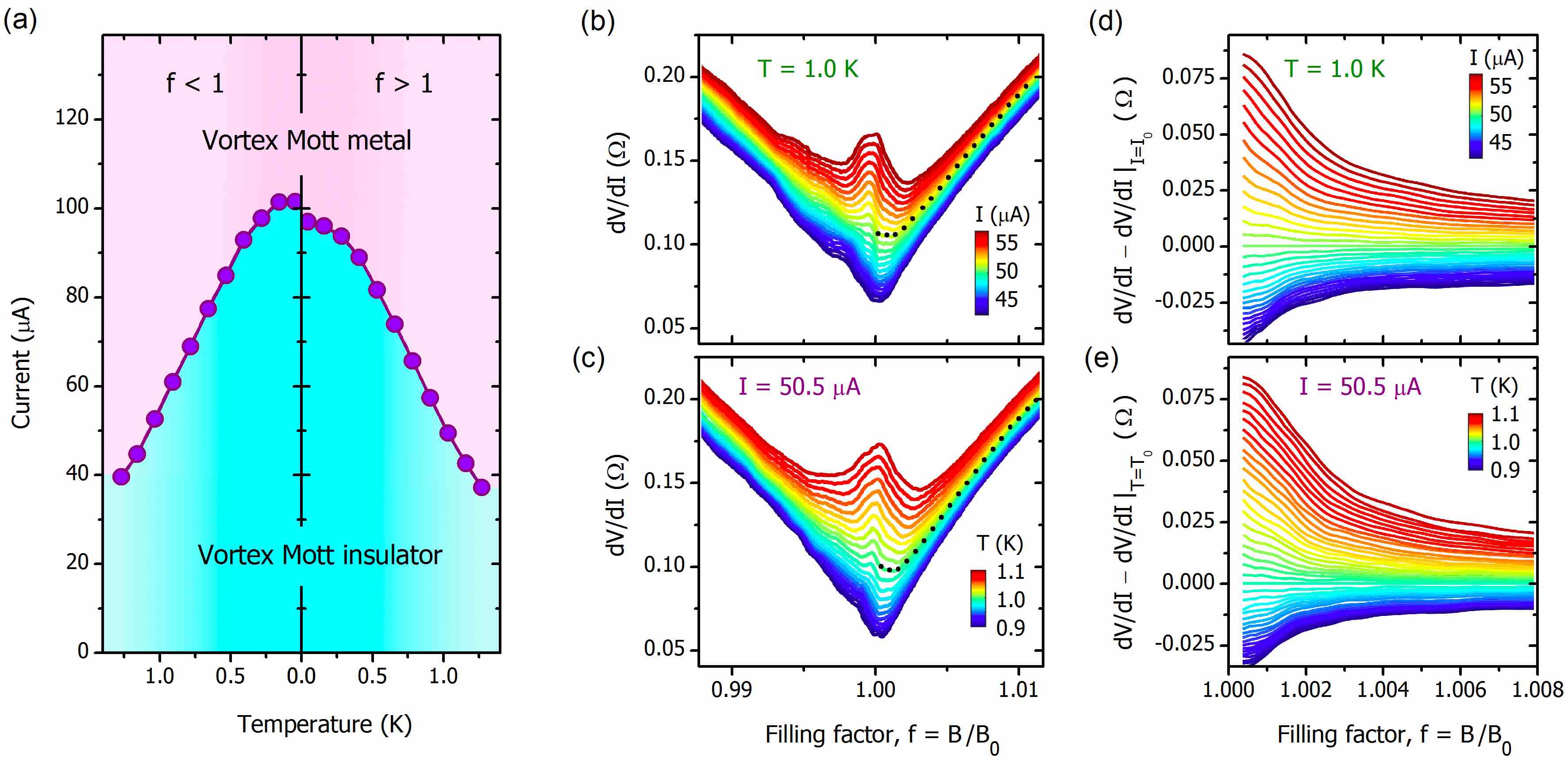}
		\caption{Vortex dynamic Mott insulator-to-metal transition.
			(a) The temperature-current phase diagram of the vortex Mott states. The left and right panels present the transition line between the insulating and metallic states at $f<1$ and $f>1$, respectively. In the former the elementary excitations are vortex holes, i.e. some of the traps lack vortices. In the latter the elementary excitations are the excess vortices i.e. some traps contain more than one vortex. 
			(b) The set of differential resistance vs. filling factor curves taken at different currents in the critical region at temperature $T=1.0$\,K. The set corresponds to current-wise crossing of the phase boundary. The currents increase from the bottom to the top, the range of currents is shown in the color legend. 
			The black dotted line is the separatrix $dV/dI|_{I=I_0}$, $I_0 = 51.0\,\mu$A for $f>1$. 
			The separatrix divides current ranges corresponding to the vortex Mott insulator (at  $I<I_0$, $dV/dI$ bend down as $f\to 1$) and vortex Mott metal (at $I>I_0$,  $dV/dI$ turning up as $f\to 1$). 
			(c) The similar set of differential resistances vs. filling factor curves, but taken at different temperatures and fixed current $I=50.5\,\mu$A. The temperature increases from the bottom to the top and corresponds to the temperature-wise crossing of the phase boundary line. 
			The black dotted line is the separatrix $dV/dI|_{T=T_0}$, $T_0 = 1.0$\,K for $f>1$. 
			(d,e) The differential magnetoresistances $dV/dI$ after subtracting the separatrices $dV/dI|_{I=I_0}$ and $dV/dI|_{T=T_0}$, respectively. The fan-like sets of curves near $f=1$ visualize the dynamic Mott transition. 
		}
		\label{Fig2}
	\end{center}
\end{figure*}
%%%%%%%%%%%%%%%%%%%%%%%%%%%%%%%%%%%%%%%%%%%%%%%%%%%%%%%%%%%%%%%%%%%%%%%%%%%%%%%%%%%%
%%%%%%%%%%%%%%%%%%%%%%%%%%%%%%%%%%%%%%%%%%%%%%%%%%%%%%%%%%%%%%%%%%%%%%%%%%%%%%%%%%%% 
%%%%%%%%%%%%%%%%%%%%%%%%%%%%%%%%%%%%%%%%%%%%%%%%%%%%%%%%%%%%%%%%%%%%%%%%%%%%%%%%%%%%%
%%%%%%%%%%%%%%%%%%%%%%%%%%%%%%%%%%%%%%%%%%%%%%%%%%%%%%%%%%%%%%%%%%%%%%%%%%%%%%%%%%%%%
\begin{figure}[b!]
	\begin{center}
		\includegraphics[width=1\linewidth]{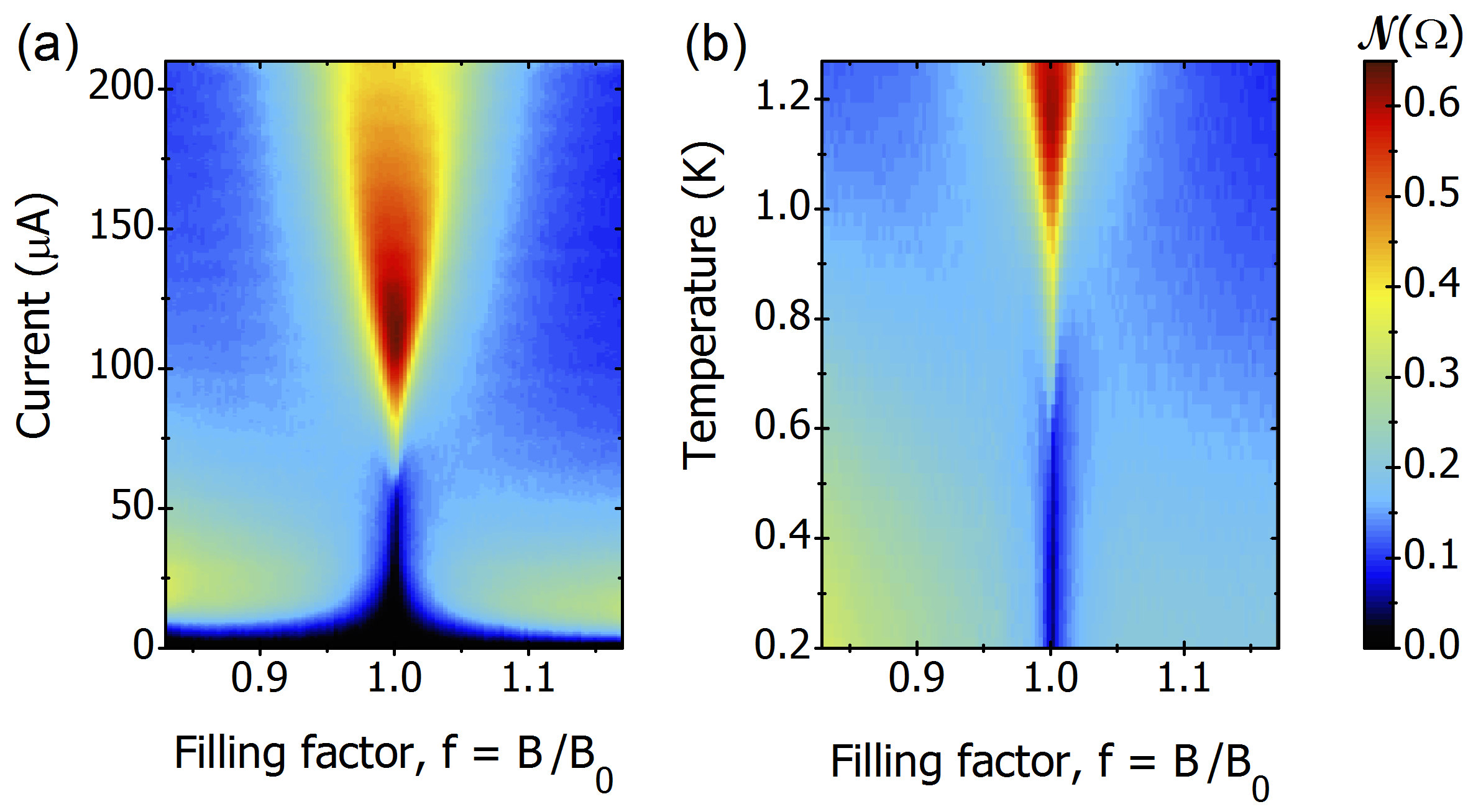}
		\caption{The critical region of the vortex dynamic Mott transition.
			The color plots of the measure of degree of nonlinearity ${\cal N}=dV/dI-V/I$ as function of the filling factor $f$ and current at $T = 1.0$\,K (a) and
			as function of $f$ and temperature at $I = 90\,\mu$A (b). The color legend is the same for both plots. 
		}
		\label{flame}
	\end{center}
\end{figure}
%%%%%%%%%%%%%%%%%%%%%%%%%%%%%%%%%%%%%%%%%%%%%%%%%%%%%%%%%
%%%%%%%%%%%%%%%%%%%%%%%%%%%%%%%%%%%%%%%%%%%%%%%%%

\section*{Critical scaling}
We start our analysis with the following question: is the observed current-driven dip-to-peak flip indeed a purely dynamic effect, or rather a mere result of the heating due to current-induced vortex motion? 
To answer it, let us consider the quantity ${\cal N}(T,B)=\mathrm{d}V/\mathrm{d}I-V/I$ that measures the degree of nonlinearity.
Figures\,\ref{flame}(a) and\,\ref{flame}(b) show the color plots of ${\cal N}(T,B)$ in coordinates $f$-$I$ and $f$-$T$, respectively.
The bright red regions (`red flames') indicate domains of strong nonlinearity that arise around $f=1$. 
Apart from the critical region near the transition the plots are predominantly blue in color. This shows that the response of the system is almost linear, $dV/dI\approx V/I$.
Since dissipation is proportional to $I\cdot V$ and the experiment is carried out at constant $I$, the dissipation is higher where $R$ is larger. Within the experimental range of currents and temperatures across the transition, the resistance grows linearly $R\propto|b|\equiv|1-f|$ upon the deviation from $f=1$. This reflects the linear increase of the
density of vortex ``holes" or the excess vortices that mediate the motion of the vortex system upon deviation from $f=1$, 
see Fig.\,6(c,d) in SM, and implies that the mobility of vortices remains nearly unchanged in our experiment. Therefore, had the nonlinearity originated from heating,
it could have only increased with increasing $|b|$. The observed effect is the opposite: the nonlinearity associated with the dip-to-peak reversal exists only in the nearest vicinity of $f=1$. 
Hence the contribution of heating effects from vortex motion is negligible 
and cannot be the origin of the observed 
dip-to-peak flip in the differential resistance.
%%%%%%%%%%%%%%%%%%%%%%%%%%%%%%%%%%%%%%%%%%%%%%%%%%%%%%%%%%%%%%%%%
%%%%%%%%%%%%%%%%%%%%%%%%%%%%%%%%%%%%%%%%%%%%%%%%%%%%%%%%%%%%%%%%%
\begin{figure*}[t!]
	\begin{center}
		\includegraphics[width=1\linewidth]{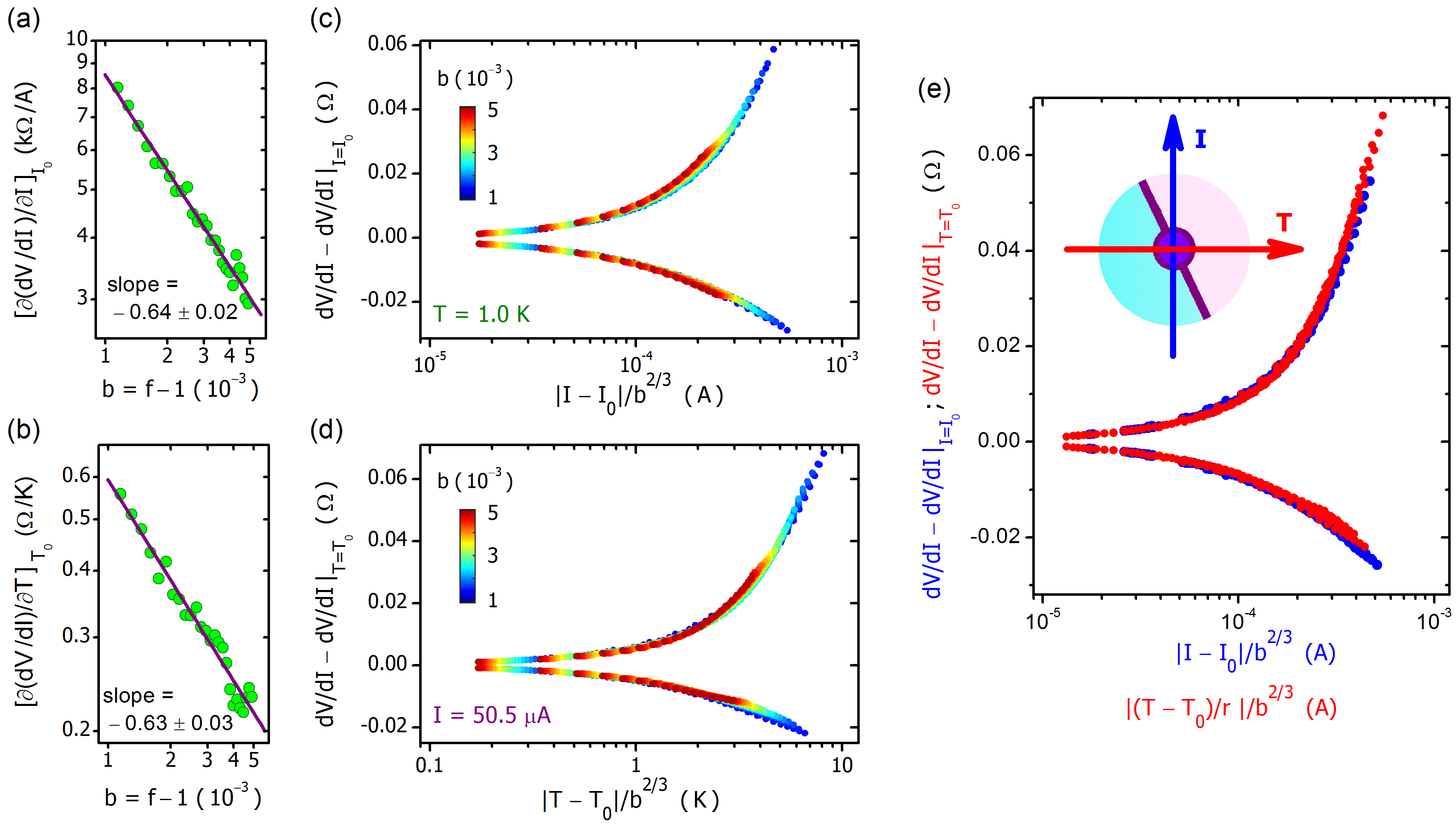}
	\end{center} %\vspace{-4mm}
	\caption{Scaling analysis of the dynamic Mott transition. 
		(a,b) The log-log plots of $[\partial(dV/dI)/\partial I]_{I_0}$ and of
		$[\partial(dV/dI)/\partial T]_{T_0}$ vs. $b$, both 
		shown by symbols. 
		The solid lines show the linear fits.
		(c) The semi-log plot of the differential magnetoresistances $dV/dI$ after subtracting the separatrix $dV/dI|_{I=I_0}$ presents the same data as Fig.\,\ref{Fig2}(d) as function of the scaling variable $|I-I_0|/b^{2/3}$.
		The perfect collapse onto two generic scaling curves for $I<I_0$ and $I>I_0$ at $\epsilon_{\rs I}=2/3$ evidences the critical behaviour of the current-driven vortex Mott transition. 
		(d)\,The semi-log plot of the differential magnetoresistances $dV/dI$ after subtracting the separatrix $dV/dI|_{T=T_0}$ presents the same data as Fig.\,\ref{Fig2}(e) as function of the scaling variable $|T-T_0|/b^{2/3}$.
		This illustrates the critical behaviour of the temperature-driven crossing of the DMT transition line. 
		(e) The plots from panels (c) (blue symbols) and (d) (red symbols) perfectly collapse on top of each other upon
		rescaling the abscissa of the panel (d) by factor $1/r$ with $r=1.5\cdot 10^4$\,K/A, evidencing 
		the identity of the ${\cal F}_{\rs I}$ and ${\cal F}_{\rs T}$ scaling functions defined by Eqs.(\ref{scalingdynamic}),(\ref{scalingthermal}). The inset shows the segment of the phase transition line. The blue and red arrows stand for current-driven and temperature-driven crossings of the transition line, respectively.
	}
	\label{criticality}
\end{figure*}
%%%%%%%%%%%%%%%%%%%%%%%%%%%%%%%%%%%%%%%%%%%%%%%%%%%%%%%%%%%%%%%%%
%%%%%%%%%%%%%%%%%%%%%%%%%%%%%%%%%%%%%%%%%%%%%%%%%%%%%%%%%%%%%%%%%

At first glance, the fact that temperature can delocalize vortices as well as current comes as no surprise.  However, closer inspection reveals that while the applied current decreases the activation barrier for vortex motion via a mere tilt of the egg-crate potential, the effect of temperature, is by far more complex. It manifests via an interplay of thermal suppression of the Josephson coupling and the smoothing of the egg-crate potential due to thermal fluctuations in vortex positions. One might thus expect quite disparate behaviors with respect the temperature and the current. Yet the detailed examination of hundreds of recorded $dV/dI$ curves versus temperature and magnetic field uncovers striking and far reaching affinity between current and temperature manifestations in the DMT critical behaviour.
The scaling analysis of the 
DMT using the representative set of $dV/dI$ curves from Fig.\,\ref{Fig2}(b,c) is shown in Fig.\,\ref{criticality}.
The benchmarks of Mott transition are the scaling relations governing the 
behaviour of $dV/dI$ in the critical region\,\cite{Imada:1998,Balents:2005,Sachdev:book,Quantum:2012,Science2015}:
\begin{eqnarray}
	\label{scalingdynamic}
	\frac{dV}{dI}(b,I,T) - \frac{dV}{dI}(b,I,T)|_{I=I_0} \propto{\cal F_{\rs I}}\left(\frac{|I - I_0|}{b^{\epsilon_{\rs I}}}\right)\,,\\
	\label{scalingthermal} 
	\frac{dV}{dI}(b,I,T) - \frac{dV}{dI}(b,I,T)|_{T=T_0} \propto{\cal F_{\rs T}}\left(\frac{|T - T_0|}{b^{\epsilon_{\rs T}}}\right),
\end{eqnarray}
where $\epsilon_{\rs I}$ and $\epsilon_{\rs T}$ are exponents describing the current- and temperature-driven critical behaviours near DMT, respectively, and  $b\equiv |f-1|$.

The formal procedure introduced in Ref.\,\cite{Hebard:1990} for determining these critical exponents is to evaluate the derivative of the dynamic resistance with respect to $I$ (or $T$) at its critical value $I_0$ ($T_0$).
Taking into account that ${\cal F}^\prime_{{\rs I},{\rs T}}(0)$ are constants, we arrive at 
\begin{equation}
[\partial(dV/dI)/\partial I]_{I_0}\propto b^{-\epsilon_{\rs I}}\,,\,\,\,\, [\partial(dV/dI)/\partial T]_{T_0}\propto b^{-\epsilon_{\rs T}}\,.
\label{derivatives}
\end{equation}
Plotting $[\partial(dV/dI)/\partial I]_{I_0}$ and $[\partial(dV/dI)/\partial T]_{T_0}$ as functions of $b$ on a log-log scale should yield straight lines with slopes equal to 
$-\epsilon_{\rs I}$ and $-\epsilon_{\rs T}$, respectively. The results of this procedure are displayed in Fig.\,\ref{criticality}(a,b).
The data are indeed the straight lines for both, current and temperature derivatives, and
the linear fit yields exponent values 
 $\epsilon_{\rs I} = 0.64 \pm 0.02$ and $\epsilon_{\rs T} = 0.63 \pm 0.03$.
Note, that the employed approach to determining scaling exponents uses only the values of the  
derivatives of the dynamic resistance at the critical point as given by Eq.\,(\ref{derivatives}).
This gives us a good starting point for the scaling analysis of the entire set of data
following Eqs.(\ref{scalingdynamic}), (\ref{scalingthermal}).
In the Figs.\,\ref{criticality}(c,d) we plot the data of Figs.\,\ref{Fig2}(b,c) as functions of the 
scaling variables $|I - I_0|/b^{\epsilon_{\rs I}}$ and $|T - T_0|/b^{\epsilon_{\rs T}}$, respectively. The collapse of the data on the single curves is excellent over two orders of magnitude of scaled abscissas for identical values of exponents
 $\epsilon_{\rs I}=2/3$ and $\epsilon_{\rs T}=2/3$. 
The same $\epsilon_{\rs I}$ for the current-driven transition was reported previously\,\cite{Science2015} for the similar proximity system, but with the distinctly different parameters. Namely, the critical temperature was significantly higher, 7.3\,K, and the island separation was smaller by factor two compared to the present case.
More results of the scaling analysis supporting the universality of the critical exponents are given in Fig.\,7 of the SM. As a next step we superimpose the scaling curves from panels (c) and (d) by
dividing the temperature abscissa by the factor $r=1.5\cdot 10^{4}$\,K/A, which on a log scale corresponds to a mere shift of the curves, see Fig.\,\ref{criticality}(e). 
The striking collapse of the isocurrent and isothermal scaling curves heralds universality of the critical scaling at the DMT.
The identity of the scaling functions 
${\cal F}_{\rs I}$ and ${\cal F}_{\rs T}$ from Eqs.\,(\ref{scalingdynamic}) and (\ref{scalingthermal}) together with the equality $\epsilon_{\rs I}=\epsilon_{\rs T}$, establishes the interchangeability of temperature and current effects in the critical region.
Finally, the collapse evidences the linear relation between the current- and temperature-induced effects and thus completely rules out the heating origin of the current-driven transition.

\section*{Discussion and theory}
We begin our theory discussion by noting that the experimental value $\epsilon_{\rs T}=2/3$ coincides with the similar exponent for the thermodynamic Mott transition 
in an electronic system\,\cite{Limelette:2003}. This implies that the thermodynamic Mott critical behavior extends onto far-from-equilibrium DMT. We conjecture that the nonequilibrium extension of the Ginzburg-Landau theory\,\cite{Chtchelk:2009} applies to thermodynamic Mott transition and that the derivation of Landau functional in Ref.\,[\onlinecite{Kotliar:2000}] can be generalized onto the DMT by including the driving current on the same footing as temperature. In the presence of the current the linear form eliminating the quadratic term in the Landau functional for the order parameter generalizes to ${\cal L}(|f-1|, T-T_0, I-I_0)\equiv \mathrm{const}_f(f-1)+\mathrm{const}_T(T-T_0)+\mathrm{const}_I(I-I_0)$.
Accordingly, the condition that ${\cal L}=0$ near the transition\,\cite{Kotliar:2000} implies that $\mathrm{const}_T(T-T_0)+\mathrm{const}_I(I-I_0)=0$ if we put $f=1$. This gives rise to $(T_0-T)/(I_0-I)=-\mathrm{const}\simeq(dT_0/dI_0)$.
Making use of the phase diagram in Fig.\,\ref{Fig2}(a), one finds at $T_0=1.0$\,K, 
$(dT_0/dI_0)=1.7\cdot 10^{4}$\,K/A in a fair agreement with the experimental rescaling factor 1.5$\cdot 10^{4}$\,K/A. 

To gain insight into the meaning of this parameter, let us recall that the energy that sets the depth of the potential well localizing vortices is estimated for a square sinusoidal egg-crate potential as $0.2E_J$\,\cite{Lobb:1983}, where the Josephson coupling of a single junction $E_J=(\hbar/2e)i_c$, $i_c=I_c/(N-1)$ is the critical current for a single junction,
$I_c$ is the critical current of the array, and $N$ is the number of rows in the array.
In this case we find that the fundamental temperature-to-current conversion ratio for a Josephson junction array $[T/I]$\,$\equiv$\,$0.2E_J/[(N-1)k_{\rs B}i_c]=1.77\cdot 10^4$\,K/A -- which nicely compares with the experimental $r=1.5\cdot 10^{4}$\,K/A. That $[T/I]\gtrsim r$ -- indicates 
that the dielectric breakdown of the Mott insulator occurs under the condition
that vortices are still pinned, which accords with our direct observation.

%%%%%%%%%%%%%%%%%%%%%%%%%%%%%%%%%%--THEORY--%%%%%%%%%%%%%%%%%%%%%%%%%%%%%%%%%%
Identical scaling functions and resulting interchangeability of current and temperature have far reaching consequences, most notably, that
dynamic critical behaviour of the Mott transition would teach us about thermodynamic criticality as well.
To construct a theory of the DMT, we first find whether vortices behave like 3${\cal D}$ or 2${\cal D}$ objects with respect to vortex line wiggling induced by thermal fluctuations, in the conditions of our experiment.  The proximity length in gold films induced by the Nb islands is expected to be $\gtrsim 100\,$\,nm (see, for example\,\cite{Mota}), which exceeds the thickness, 40\,nm of the gold substrate. Therefore, one expects that Nb islands induce superconductivity throughout the gold substrate. Therefore, the system can be viewed as a superconducting film with periodically modulated thickness, the thinnest volleys corresponding to inter-island areas.  To check whether a vortex can be viewed as a flexible 3${\cal D}$ string, one has to verify that the longitudinal size of the thermal vortex fluctuation, $\ell_{\rs T}$, fits within the gold film thickness. One estimates $\ell_{\rs T}\sim u_{\rs T}^2\varepsilon/k_{\rs B}T$\,\cite{Nelson}, where $u_{\rs T}$ is the average lateral thermal fluctuation of the vortex line and $\varepsilon$ is its linear tension. The maximal possible $u_{\rs T}^{\mathrm{max}}$, such that the energy of the thermal fluctuations would not exceed the elastic interactions with other vortices localizing the test vortex within the potential well (i.e. that thermal fluctuations do not melt vortex lattice), is given by $u_{\rs T}^{\mathrm{max}}\simeq c_{\rs L}a$,\,\cite{Vinokur:1998,Blatter:1994} where $a\approx 200$\,nm at $f=1$ is the equilibrium vortex spacing. Making use of the relation $\varepsilon a\simeq 6T_m\approx 6T_0$, $T_m$ being the vortex lattice  melting temperature,   
and recalling that for the vortex lattice $c_{\rs L}=0.16$\,\cite{Vinokur:1998}, one finds that under the conditions of the experiment the longitudinal fluctuations do not exceed $\ell_{\rs T}\simeq 30$\,nm. This means that vortices can bend exercising 3${\cal D}$ thermal fluctuations and hence the 3${\cal D}\to2{\cal D}$ quantum mechanical mapping applies.
In this mapping, the thermally activated motion of a 3${\cal D}$ vortex over the energy barrier corresponds to the quantum tunneling of a 2${\cal D}$ quantum particle across the same barrier.
This enables us to describe the decay of the vortex Mott insulator as the electric field-driven Landau-Zener-Schwinger (LZS) tunneling of a charged quantum particle across the Mott gap, $\Delta$,\,\cite{Oka:2003,Oka:2010} and 
construct a quantitative non-Hermitean LZS theory of the critical behaviour of the DMT\,\cite{tripathi:2016}.
We find that in the presence of dissipation the applied electric
field $F$ generates an imaginary field $\chi(F)$. The Hamiltonian becomes non-Hermitean while retaining its invariance under 
the combined parity reflection and time reversal (${\cal PT}$) transformation.
The Mott gap is defined as $\Delta\equiv E_1-E_0$, where $E_0$ and $E_1$ are the energies of the ground, $|0\rangle$, and the first excited, $|1\rangle$, eigenstates.
Remarkably, within a framework of non-Hermitean LZS, the very definition of $\Delta$ that prohibits the introduction of the standard order parameter for Mott insulator, ceases to be 
a stumbling block for a theory, but becomes a stepping stone enabling a description of the Mott insulator's decay as the probability of the LZS tunneling, $P=|\langle 0|1\rangle|^2$, across the gap. The DMT occurs at the field where $P$ becomes unity.
At the critical value $\chi_0=\chi(F_0)$, where $F_0$ is the field of the dielectric breakdown,
the eigenvalues $E_0$ and $E_1$ merge and the Mott gap collapses to zero (see Appendix for the details of calculation).
Simultaneously, at this bifurcation point the ground state loses its  ${\cal PT}$ symmetry.
We thus identify DMT as the ${\cal PT}$ symmetry-breaking phase transition.
Analyzing the spectrum behavior near the bifurcation point, we find the critical collapse of the Mott gap as
$\Delta\propto (F_0-F)^{1/2}$ leading to $P\sim\exp(-2\gamma)$ with the effective action $\gamma\propto(F_0-F)^{3/2}/F$.
Upon returning to the vortex system via the reversed quantum mapping, i.e. substituting $\hbar$ by temperature $T$ and the field $F$ by the current $I$, the probability of the decay of the vortex Mott insulating state assumes the thermally activated form 
$P\propto\exp[-A(I_0-I)^{3/2}/T]$ ($A$ is the constant to ensure the correct dimensionality) 
with the activation barrier that scales as $(I_0-I)^{3/2}$ near the DMT.
This leads directly to the critical exponent $\epsilon_{\rs I}=2/3$ (see Methods) and is exactly what our experiment shows.

In the opposite limit of very thin superconducting system such that $\ell_{\rs T}$ exceeded the thickness of the proximity-induced superconducting film, the vortices are effectively two-dimensional and quantum mapping would not apply. However, as we now show, the non-Hermitian description of the DMT, leading to the $(I_0-I)^{3/2}$ scaling, holds even in this case. To see that, note that in the vicinity of the commensurability point, $f=1$, the transport properties of a near-commensurate vortex system are governed by the density of the excess/deficit excitations over its commensurate value. We thus introduce a classical field $\Psi({\bf x},t)$ describing the excess vortices (or vortex holes). The fluctuating part of the vortex system free energy is then given by the $2{\cal D}$ Ginzburg-Landau functional 
\begin{equation}
{\cal F} %&
 =\int d^{2}x\left[D|\nabla\Psi|^{2}+m^{2}|\Psi|^{2}+u|\Psi|^{4}\right]\,,\label{eq:LG-model}
\end{equation}
where $D$ is the stiffness of the excess vortices system, and $m$
and $u$ are respectively the mass and interaction parameters that
govern the mean-field transition.
Although the vortex fields are not intrinsically dynamic, they are subject to temporal
fluctuations due to coupling to the Ohmic environment of the metallic vortex cores. 
This results in the overdamped equation of motion,
\begin{align}
\frac{\partial\Psi}{\partial t} & +\rho\frac{\delta {\cal F}}{\delta\Psi^{*}}=0,\label{eq:eq-motion}
\end{align}
where $\rho$ represents viscous damping of the vortex motion and is phenomenologically
proportional to the (charge) resistivity. 
Performing gauge transformation
to turn the vector potential into the scalar one, we can recast
Eq.\,\eqref{eq:eq-motion} into the form
\begin{align}
\frac{\partial\Psi}{\partial t}-i(I/\rho)x\Psi & =D\nabla^{2}\Psi-m^{2}\Psi-2u|\Psi|^{2}\Psi\,,\label{eq:sch-eqn}
\end{align}
where $I$ is the applied imaginary current driving vortices.
This equation is formally identical to a nonlinear Schr\"odinger equation in
Euclidean time for 2D interacting particles subject to an imaginary electric field.
In the vicinity of the transition one can neglect the nonlinear term, and again exercise the machinery of the
LZS tunneling of a charged quantum particle across the Mott gap described above and find $(I_0-I)^{3/2}$ scaling near the DMT.

Note finally that our $\cal{PT}$ symmetry-based description of the DMT rests on the general properties of non-Hermitian quantum mechanics rather than on specifics characteristic to Mott systems.  
Therefore, our approach applies to a broad class of phenomena well beyond the immediate context of the Mott physics and
provides a universal tool for analytical description of out-of-equilibrium phase transitions and instabilities in open quantum dissipative systems. 
%to a  of open quantum dissipative many-body strongly correlated systems 
The general principle is that if at small drives the system's non-Hermitian Hamiltonian is endowed with the $\cal{PT}$ symmetry, the out-of-equilibrium phase transition manifests as a $\cal{PT}$ symmetry-breaking at the corresponding threshold
value of the driving field.

\section*{Appendix A: Derivation of the Mott gap collapse}
The vortices frozen into the minima of the egg-crate potential near $f=1$ are described by the Hubbard model\,\cite{Oka:2003,Oka:2010}. The Mott gap, $\Delta(\Psi(t))=E_{1}-E_{0}$, is defined as the difference of energies of the first excited state $E_1$ and the ground state $E_0$ of the Hubbard Hamiltonian, with $\Psi(t)$ being the time-dependent gauge field describing the combined effects of the 
 applied field $F$ and dissipation. The rate of the decay of the Mott insulating state is given by the LSZ probability of the $|0\rangle\to|1\rangle$ transitions, $P\equiv|\langle 0|1\rangle|^2\sim \exp(-2\gamma)$.
 The reduced action is
 given by the Landau-Dykhne formula\,\cite{Dykhne:1962},
 $\gamma =(1/\hbar)\,\text{Im}\int dt\, [E_{1}(\Psi(t))-E_{0}(\Psi(t))]$,
 with $\Psi=Ft+\mathrm{i}\chi$. The
 imaginary part of the field, $\chi$, which arises as a result of combined action of driving field and dissipation, increases monotonously  with the applied field $F$ and renormalizes the Mott gap $\Delta$ to zero at the Mott transition. 
The dissipation, i.e. energy relaxation, makes the quantum amplitudes along and opposite to the applied field unequal, the
 difference being quantified by the factor $e^{2\chi}$, see Ref.\,\cite{Nelson:1998}. 
 One can show analogously to\,\cite{Chtchelk:2009} that integrating out the thermal bath degrees of freedom generates the imaginary potential $\chi\sim iF$ near the critical point.
 The Mott gap closes at the critical field $F=F_0$ which maps onto the critical
  point $\Psi_c$ in the complex $\Psi$-plane.
 In terms of the non-Hermitian Hamiltonian, merging of $E_0$ and $E_1$ and closing the corresponding spectral gap marks the 
  ${\cal PT}$ symmetry breaking transition. 
  At this point the eigenstate $|0\rangle$ loses its ${\cal PT}$ symmetry and $E_0$ simultaneously acquires the imaginary part, i.e. the energy spectrum ceases to be real\,\cite{Bender}.  The DMT is then identified
 with the ${\cal PT}$ symmetry-breaking phase transition. 
 The Mott insulating state 
 corresponds to the regime of unbroken ${\cal PT}$ symmetry with the real energy spectrum\,\cite{tripathi:2016}.
 The exact field-dependence of the imaginary part of the gauge field, $\chi(F)$, depends on microscopic details, but for analysis of the critical behaviour it suffices to know that $\chi(F)$ is a well-behaved function of $F$ near the critical field $F_0$.
 Expanding around the critical value $\chi_{0}\equiv\chi(F_0)$, where
 $\Delta(F_0)=0$, yields %\textcolor{blue} {in our case} 
 $\gamma\approx(1/\hbar)\,\int_{\chi}^{\chi_0} d\chi'\, \Delta(\chi')/|d\Psi/dt|=(1/F\hbar)\int_{\chi}^{\chi_0} d\chi'\, \Delta(\chi')\equiv (F_{\text{th}}/F) \propto(F_0-F)^{3/2}/F$ (see SI).
 Far below the transition $\gamma$ reduces to the standard Landau-Zener formula\,\cite{Oka:2003,Oka:2010}, where the threshold field $F_{\text{th}}$ is related to the Mott gap $\Delta\sim |U - U_c|$ 
 as $F_{\text{th}} \sim \Delta^2$. 
 Here $U$ is interaction strength, and $U_c$ is the critical interaction at which the Mott transition takes place. To relate the critical exponent for the collapse of the Mott gap with $\epsilon_I$, we recall that in
  the system of superconducting vortices, their
 interaction strength is controlled by the
 vortex density, which is proportional to
 the external magnetic field, hence $|U - U_c|\sim|b|$.
 Replacing the field $F$ by the current $I$ according to the quantum mapping recipe,
 the energy gap collapses non-analytically as $\Delta\sim|\chi_0 - \chi|^{1/2}$, upon approaching to the Mott transition, $I\to I_0$, see\,Ref.\,[\onlinecite{Chtchelk:2012}]. 
 Accordingly, the threshold current $I_{\text{th}}$ ($\gamma=I_{\text{th}}/I$) scales as $|I_0 - I|^{3/2}$ since $\Delta\sim|\chi_0 - \chi|^{1/2}$.  Since near the DMT, the kinetic energy gained by a vortex in nearest-neighbour hopping down an effective field $I_{\text{th}}$ 
scales as  $E_J$ (corresponding to the strength of the Coulomb repulsion in quantum particle representation), the universal scaling function is a homogeneous function of $|I-I_0|^{3/2}/b,$ i.e., $\epsilon_I = 2/3$. 
In the SM, we provide details of the scaling analysis and discuss the relation between $\epsilon_I$ and the standard critical exponents $z$ and $\nu.$

\subsection{Acknowledgments}
 We thank Frank Roesthuis and Dick Veldhuis for help
  and support during the experiments and I. Aleiner, B. Altshuler, G. Kotliar, and
  A. Millis for illuminating discussions. Work was supported by the Dutch FOM
  and NWO foundations, the Italian Ministry for Education and Research, the
  Russian Science Foundation (project No 14-22-00143), the Ministry of Education
  and Science of the Russian Federation, and by the U.S. Department
  of Energy, Office of Science, Materials Sciences and Engineering Division (A.G. and V.M.V.);
  V.T. was supported through Materials Theory Institute at ANL, the University
  of Chicago Center in Delhi and a DST (India) Swarnajayanti grant (no.
  DST/SJF/PSA-0212012-13). T.I.B. acknowledges financial support from the
  Alexander von Humboldt Foundation and from the Consejería de Educación, Cultura y Deporte (Comunidad de Madrid) through the talent attraction program, Ref. 2016-T3/IND-1839.

%\newpage
%\onecolumn
\section{Supplementary Materials}

\subsection{Experimental methods}
The device was fabricated on a SiO$_2$-covered Si substrate. The 40\,nm thick gold square was patterned using photolithography and sputter deposition. On top of the Au film the square array of 270-by-270 Nb dots was deposited using standard e-beam lithography and DC sputtering. The Nb layer has a thickness of 45\,nm. The dots have diameters of $142\pm 5$\,nm, the center-to-center distance between two adjacent dots is 250\,nm. On  either side of the array a Nb crossbar was patterned to ensure that the current is injected homogeneously into the array. Figure S1 shows the temperature dependence of the resistance near the superconducting transition $T_c$. Transport measurements were performed in a 3He/4He dilution refrigerator. 

\begin{figure}[h!]
	\begin{center}
		\includegraphics[width=0.9\linewidth]{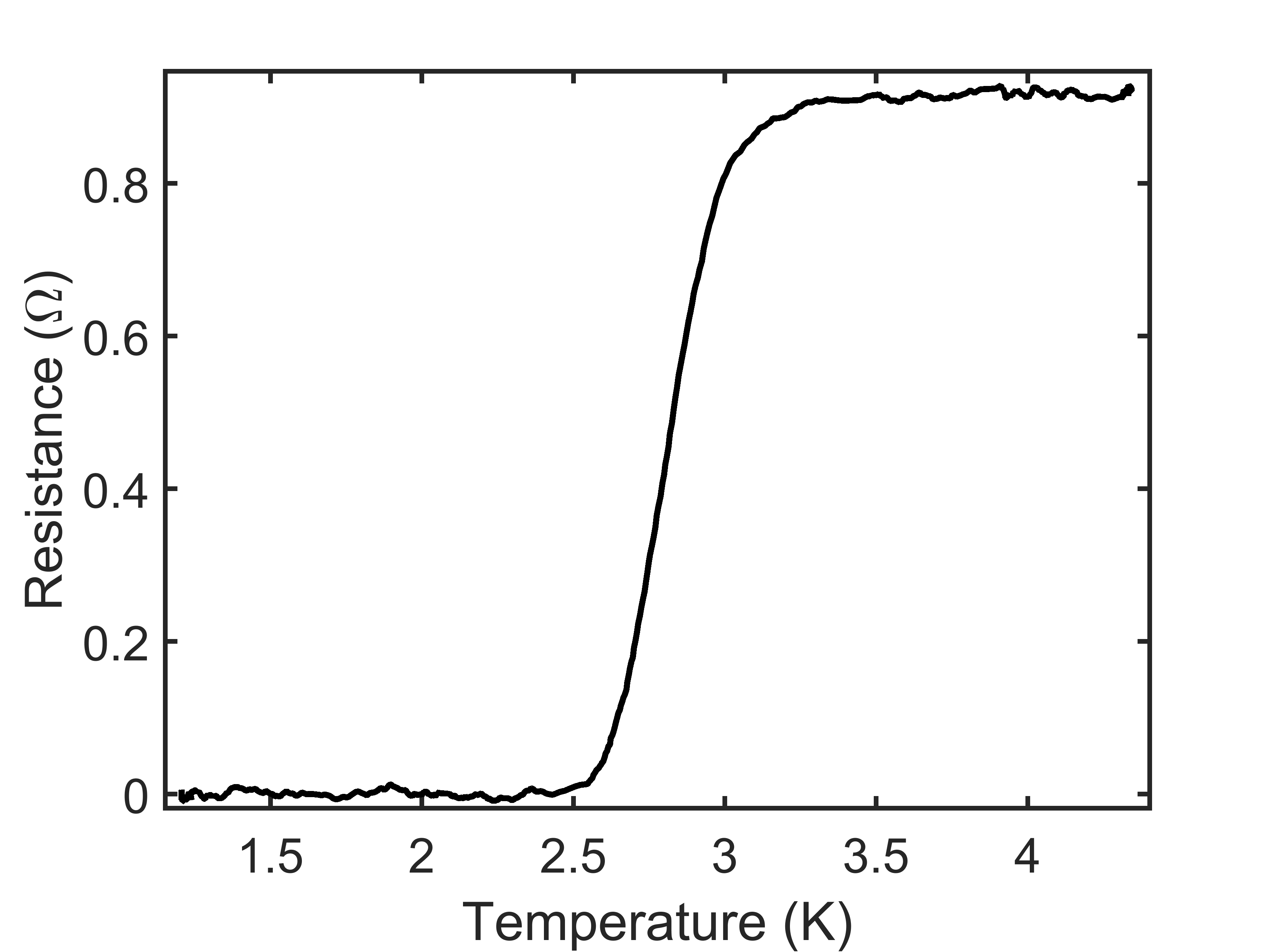} %\vspace{-3mm}
		\caption{{\bf Superconducting transition.} 
			The superconducting transition temperature of the array, determined as the midpoint of the temperature resistance curve, is $T_c = 2.7$\,K, which is 6.6\,K lower than that of bulk Nb ($T_{c0} = 9.3$\,K)
		}
		\label{Sup1}
	\end{center}
\end{figure}

Two sets of measurements were performed in the configuration where the magnetic field was perpendicular to the plane of the array. 
In the first set of measurements the $V(I)$ curves were taken at a constant temperature and at a constant magnetic field. 
This was done for multiple values of the magnetic field (between 0\,mT and 38\,mT in 0.1\,mT steps) and the temperature (between 70\,mK and 1.27\,K in 25\,mK steps). This results in the data $V(I,B,T)$, from which the differential resistance was obtained by taking a numerical derivative with respect to the current. 

To carry out the detailed scaling analysis, the high-resolution data were taken using the standard lock-in measurement technique with the AC excitation current $1\,\mu$A. 
The data presented in Fig.\,2(b,d) and Fig.\,6\,a,c are taken at fixed temperature $T=1.0$\,K. The current and magnetic field were swept over the ranges from $43\,\mu$A to $57\,\mu$A (with $0.5\,\mu$A steps) and from 32.8\,mT to 33.6\,mT (with $5\,\mu$T steps), respectively. 
The data shown in Fig.\,2(c,e) and Fig.\,6(b,d) were collected at  the fixed current $I = 50.5$\,$\mu$A while the magnetic field was swept from 32.8\,mT to 33.6\,mT with the step $5\,\mu$T and the temperature increased from 0.900\,K to 1.095\,K with the 5\,mK steps.

\subsection{Differential resistance $dV/dI$ and resistance $R=V/I$ }
Figure 6 juxtaposes the differential resistance $dV/dI$ and resistance $R=V/I$ as functions of filling factor $f$ inferred from the same set of the current-voltage characteristics. Plots in the panels \textbf{a} and \textbf{b} reproduce Fig.\,2(b,c) displaying the evolution of $dV/dI$ upon increasing $I$ (at constant $T$) and $T$ (at constant $I$), respectively. The dip to peak reversal at $f=1$ signals the crossing of the dynamic Mott transition line. Panels \textbf{c} and \textbf{d} show that the corresponding resistances maintain pronounced dips at $f=1$.  As the measurements are carried out under constant currents, the plots for the resistances 
coincide up to the numerical factor, $I^2$, with the plots for the dissipated power. 
Therefore, the dissipated power is minimal at $f=1$.

\begin{figure*}[t!]
	\begin{center}
		\includegraphics[width=0.52\linewidth]{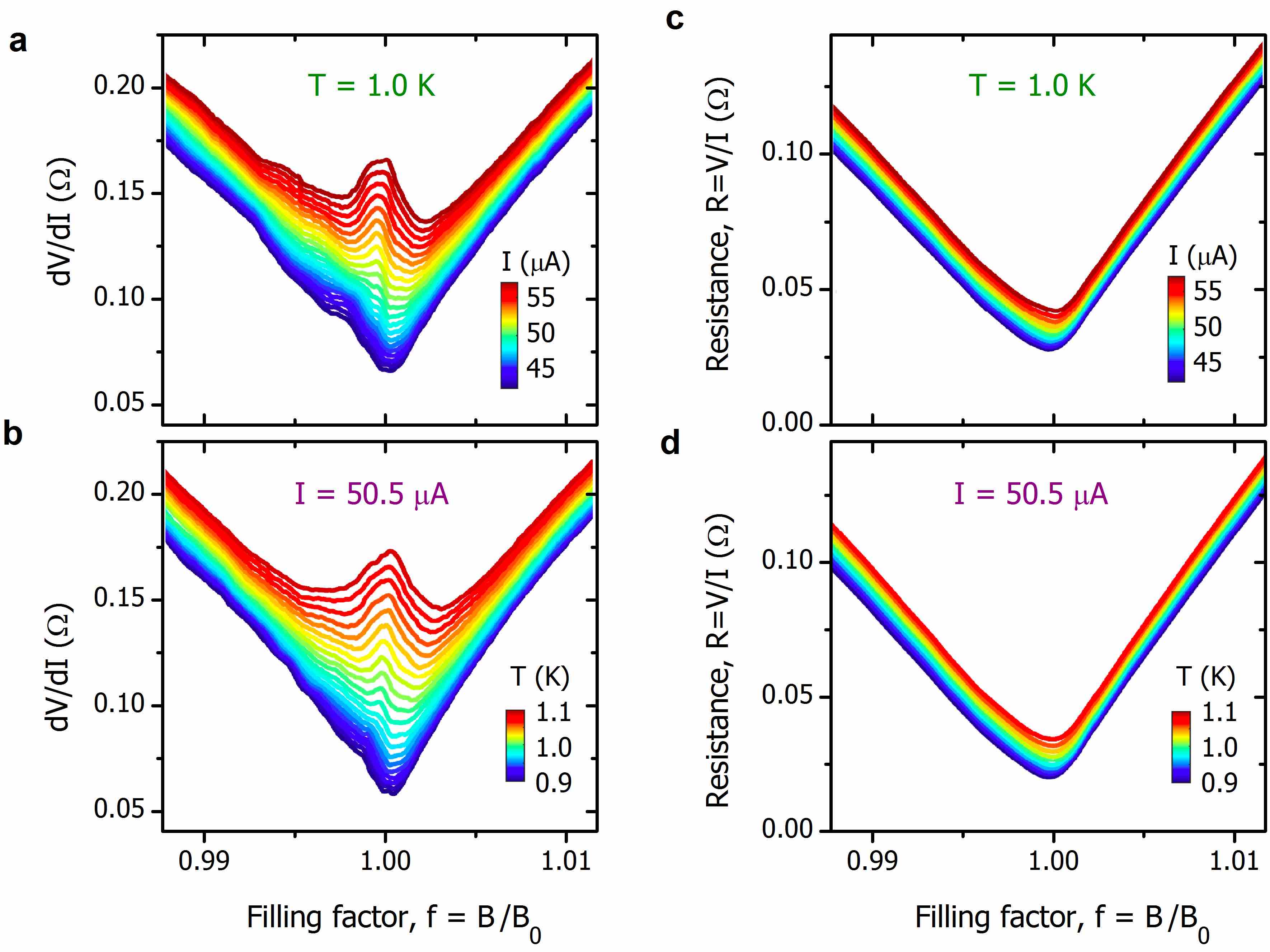}
		\caption{{\bf Vortex Mott insulator-to-metal transition.} 
			\textbf{a,c} Differential magnetoresistance $dV/dI$ and resistance respectively, taken at different currents increasing from bottom to top as a function of the filling factor $f$ in the vicinity of the commensurate value $f=1$ at $T=1.0\,$K. 
			\textbf{b,d} Differential resistance $dV/dI$ and resistance $R=V/I$, respectively, taken at different temperatures increasing from bottom to top as a function of the filling factor $f$ in the vicinity of the commensurate value $f=1$ at $I = 50.5\,\mu$A. 
		}
		\label{Sup2}
	\end{center}
\end{figure*}
\begin{figure*}[t!]
	\begin{center}
		\includegraphics[width=0.87\linewidth]{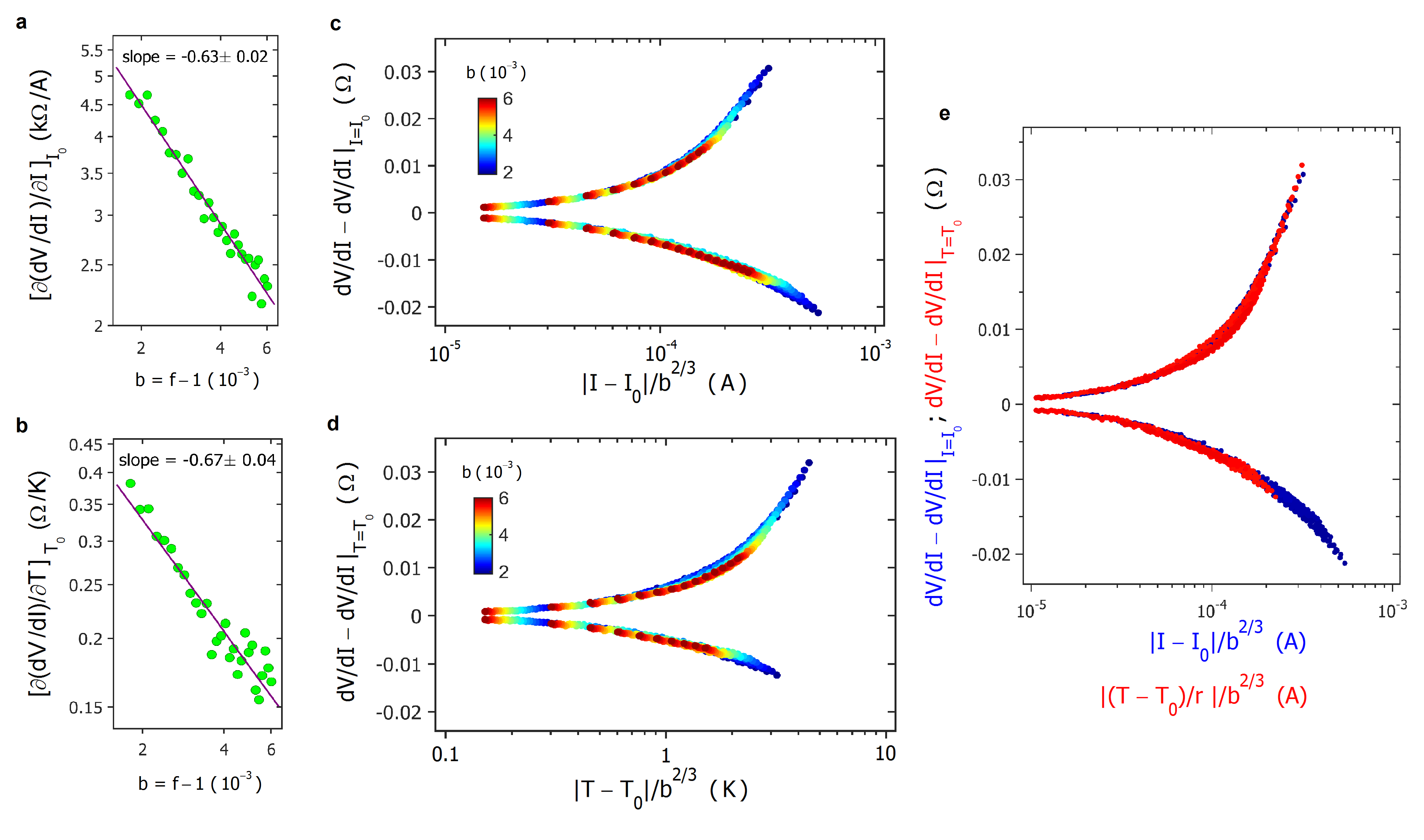}
		\caption{\textbf{Scaling analysis of the dynamic Mott transition around $T = 0.75$\,K.}  
			\textbf{a,b} The log-log plots of $[\partial(dV/dI)/\partial I]_{I_0}$ and of
			$[\partial(dV/dI)/\partial T]_{T_0}$ vs. $b$, both 
			shown by symbols.  The current driven data (\textbf{a,c}) is taken at $T = 0.75$\,K with $I_0 = 67\,\mu$A. The thermally driven data (\textbf{b,d}) is taken at $I = 68.5\,\mu$A with $T_0 = 0.73$\,K.
			\textbf{c,} The semi-log plot of the differential magnetoresistances $dV/dI$ after subtracting the separatrix $dV/dI|_{I=I_0}$ as function of the scaling variable $|I-I_0|/b^{2/3}$.
			The perfect collapse onto two generic scaling curves for $I<I_0$ and $I>I_0$ at $\epsilon_{\rs I}=2/3$ evidences the critical behaviour of the current-driven vortex Mott transition. 
			\textbf{d,}\,The semi-log plot of the differential magnetoresistances $dV/dI$ after subtracting the separatrix $dV/dI|_{T=T_0}$ as function of the scaling variable $|T-T_0|/b^{2/3}$.
			This illustrates the critical behaviour of the temperature-driven crossing of the DMT transition line. 
			\textbf{e,} The plots from panels \textbf{c} (blue symbols) and \textbf{d} (red symbols) perfectly collapse on top of each other upon
			rescaling the abscissa of the panel \textbf{d} by factor $1/r$ with $r=1.42\cdot 10^4$\,K/A. This value of $r$ is close to the value of $r=1.5\cdot 10^4$\,K/A found around $T = 1$\,K.
		}
		\label{Sup3}
	\end{center}
\end{figure*}

\subsection{Scaling analysis at $T=0.75$\,K and $I=68.5\,\mu$A}
Figure 7 shows scaling analysis around $T = 0.75\,$K. The scaling exponents $\epsilon_{\rs I}=0.63$ and $\epsilon_{\rs T}=0.67$ are found, from which we conclude that $\epsilon = 2/3$. The measured data has the same stepsizes and resolution as the data around $T = 1.0\,$K described above. This data supports the finding shown in Fig.4 of the main text.

\newpage

\subsection{Mott transitions in a vortex lattice system}
A vortex system in the presence of a finite current is equivalent
to a 2D bosonic Coulomb gas in the presence of an electric field.
Vortices have a bare mass proportional to the electrostatic charging
energy $E_{c}$ and the Coulomb interaction scale is determined by
the Josephson energy $E_{J}.$ In the presence of a periodic potential generated by the 
proximity array (analogous to lattice matrix for charged bosons), the bare vortex mass
transforms into the band
mass, which, at commensurate fillings, can become considerably larger than the bare mass.
The latter promotes vortex localization and the formation of a vortex Mott insulator. 
Careful numerical studies
\cite{ceperley} of a 2D bosonic Coulomb gas (not on a lattice) show
that the Mott transition takes place at $r_{s}\approx12,$ where $r_{s}$
is the ratio of intervortex separation and the vortex Bohr radius.
To the best of our knowledge, there is no comparable study of the 2D bosonic Coulomb
gas on the lattice. 

What has been studied in detail is a variety of bosonic Hubbard models on 2D lattices that exhibit Mott-insulator/superfluid
transitions as a function of vortex chemical potential $\mu$ and
hopping energy $t$ measured in terms of the local repulsion $U$.
Thus to be able to utilize a rich lore of the Hubbard model machinery and yet preserve the important features of the long-range vortex-vortex interaction, we adopt
a Bose-Hubbard model with hard-core repulsion
and finite nearest and next-nearest neighbour repulsive interactions
(we call these $V_{1}$ and $V_{2}$, respectively): 
\begin{align}
	H & =-t\sum_{\langle ij\rangle}(b_{i}^{\dagger}b_{j}+\text{h.c.})+V_{1}\sum_{\langle ij\rangle}n_{i}n_{j}+V_{2}\sum_{\langle\langle ij\rangle\rangle}n_{i}n_{j}\,, %\qquad n_{i}=0,1.
	\label{eq:model-hardcore}
\end{align}
where $n_{i}=0,1$.
Here $b_{i}^{\dagger}$ creates a boson at site $i$, $n_{i}=b_{i}^{\dagger}b_{i}$
is the boson number at site $i.$ Such long-range interactions open
the further possibility of spatial order of the bosons with or without
underlying superfluidity. Tuning the magnetic field changes the vortex
chemical potential as well as the inter-vortex interaction strength.
This corresponds to the trajectory in the $\mu/V_{1}$
vs $t/V_{1}$ phase diagram. The hard-core constraint makes the model
equivalent to an $XXZ$ antiferromagnet where $t$ maps to $J_{x}/2$
and $V_{1}$ maps to $J_{z}$ etc. Half-filling corresponds to the
zero magnetization sector. The chemical potential is equivalent to
the applied magnetic field $B=(\mu-zV_{1}/2)$. 

Detailed numerical studies are available for such a $V_{1}-V_{2}$
hard core boson model on a square lattice \cite{hebert}. Near $f_{c}=1,$
the Mott insulating phase consists of one vortex per site, and the
vortex Mott insulator - superfluid quantum phase transition is mean-field
like (except perhaps at the tip of the Mott lobe). In this work, we have
focused on this regime.

Alongside with the magnetic field that acts as a chemical potential for
vortices and, therefore, controls the vortex Mott transition by
varying the relative strength of vortex kinetic and potential energies,
another parameter that tunes the quantum vortex
Mott transition, is the current $I$. The current for vortices plays therole of the electric field which induces the dynamic
Mott transitions in electronic Mott insulators.

\subsection{Scaling analysis}
Near the quantum critical point $(f_{c},\, I_{0})$ corresponding
to the vortex Mott transition, the dynamic resistance $dV/dI$ is
expected to show the scaling behavior, 
\begin{align}
	\frac{dV}{dI} & =|\delta|^{p}\Phi\left(\frac{T}{|\delta|^{z\nu}},\frac{|I-I_{0}|}{|b|^{\epsilon_{I}}}\right).\label{eq:QPT-scaling-SI}
\end{align}
Here $\Phi$ is the universal scaling function depending on the
universality class of the phase transition and not on the microscopic
details of the model. The parameter $\delta=g-g_{c}$ measures the distance from
the quantum critical point $g_{c}$ in the parameter space and is itself a function
of the parameters $f_{c},\, I_{0}$, $\nu$ is the scaling exponent
for the length, and $z$ is the dynamical scaling exponent. The exponent
$\epsilon_I$ describes the relative scaling with respect to the current and the
filling fraction respectively, $p$ is the scaling exponent for the dynamic resistance,
and $b=f-f_{c}$. Before turning to estimates for $\epsilon_I,$ we discuss the 
effect of temperature on the Mott transition.

For certain universality classes of quantum phase transitions, the
scaling function $\Phi$ shows a singular behavior at finite temperatures
$T_{0}(f_{c},\, I),$ which corresponds to a classical phase transition.
The transition temperature $T_{0}$ is a function of $f$
and $I$ and its scaling with these parameters is governed by the
underlying quantum critical theory. Within a small range of temperature,
current and filling such that $|T-T_{0}(f_{c},I)|/T_{c}\ll1,$ the
dynamic resistance will have a scaling form 
\begin{align}
	\frac{dV}{dI} & \sim{\cal F}\left(\frac{|T-T_{0}|}{|b|^{\epsilon_T}},\frac{|T-T_{0}|}{|I-I_{0}|^{\Delta_{I}}}\,,\ldots\right),\label{eq:Cl-scaling-SI}
\end{align}
where $\epsilon_T$ and $\Delta_{I}$ are scaling
exponents corresponding to the finite temperature transition.
Away from the quantum critical point where the relative scaling of
the temperature and $I-I_{0}$ may be nonanalytic, we use the Taylor
expansion of the critical temperature 
$$
T_{c}(f_{c},I)\approx T_{c}(f_{c},I_{0})+(I-I_{0})T_{c}'(f_{c},I_{0})+\ldots\,\,.
$$
Then, together with Eq.\,(\ref{eq:Cl-scaling-SI}), we get the scaling
function ${\cal F}(|I-I_{0}|/|b|^{\epsilon_{T}}),$ and $\Delta_{I}=1.$
The exponent $\epsilon_T$ may be estimated as follows. In\,\cite{kotliar}
the critical behaviour of the classical Mott transition in a half-filled
Hubbard model has been studied in terms of the order parameter that
corresponds to the fraction of doubly-occupied sites. The Coulomb
correlation $U$ can then be naturally regarded as a conjugate field
that couples linearly to the order parameter. As a result, the order
parameter scales as ${\cal O}\sim(U-U_{c})^{1/\delta_T},$
where $\delta_T$ is the critical exponent associated with
scaling of the order parameter with the conjugate field. On the other
hand, as a function of temperature (or current), the order parameter
behaves as ${\cal O}\sim(T_{0}-T)^{\beta_T}.$ Since $|b|$
tunes the distance from $f_{c}$ and therefore the relative strength
of the Mott repulsion, we posit that $|b|\leftrightarrow|U-U_{c}|.$
It then follows that $\epsilon_T = 1/\beta_T\delta_T$
for the vortex Mott transition. Near $f_{c}=1,$ the Mott transition is of the mean-field type, so 
we use $\delta_T=3$ and $\beta_T=1/2$
to arrive at $\epsilon_T=2/3.$ This is remarkably consistent
with our finite temperature and finite current data. 

\subsection{Dielectric breakdown in dissipative Mott insulators\label{sec:Dielectric-breakdown}}

The driving field promotes conduction in a dissipative gapped system in two ways: (i) through generation of free particle-hole
pairs by the Landau-Zener mechanism while keeping the gap magnitude
fixed, and (ii) by renormalization of the energy gap (mass) which affects the Landau-Zener tunneling probability. 
Consider an interacting quantum system in its ground state $|0\rangle$ separated from the lowest excited state
$|1\rangle$ by the spectral gap $\Delta$. 
If the driving field is applied adiabatically,
the probability $P=|\langle0|1\rangle|^{2}\sim e^{-2\gamma}$
for the system to transit to the excited state is given by the
Landau-Dykhne formula 
%(see Ref.\cite{oka} for example), 
\begin{align}
	\gamma & =\text{Im}\int_{-\infty}^{\infty}dt\,[E_{1}(\Psi(t))-E_{0}(\Psi(t))],\label{eq:oka1}
\end{align}
where $\Psi$ denotes a time-dependent phase factor related to the
driving field $F,$ $E_{0}$ is the ground state energy (parametrically
dependent on $\Psi$) and $E_{1}-E_{0}\equiv\Delta$. 
For an electron hopping along a constant electric field, we choose the gauge where
the driving field is the time derivative of
the vector potential, hence $\Psi=Ft$ is the Aharanov-Bohm phase
acquired for a nearest-neighbour hop. Then changing variables we replace 
the integral over time with the integral over complex 
$\Psi=Ft\pm i\chi$ and deform contour in the complex $\Psi$-plane. 
The imaginary part of the integral
over $\Psi$ comes from the degeneracy point $i\chi_{c}$ in the complex
$\Psi$-plane where the gap closes. Assuming no other singularities,
we deform the $\Psi$ contour to the imaginary axis and obtain 
\begin{align}
	\gamma & =\frac{1}{F}\text{Re}\int_{\chi}^{\chi_{c}}d\chi'[E_{1}(\chi')-E_{0}(\chi')],\label{eq:oka2}
\end{align}
for the Landau-Zener tunnelling factor. In non-dissipative models, including in strongly-correlated models such
as the half-filled Hubbard chain~\cite{oka},
Eq.\,(\ref{eq:oka1}) reduces to the well-known Landau-Zener result $\gamma\sim\Delta^{2}/vF\equiv F_{\text{th}}/F,$
where $v=|d\Delta/dt|/F$ denotes the ``velocity'' of the mutual approach of
the two levels as $\Psi$ is varied, and is assumed to be a constant
(i.e. independent of $\Delta$) in the usual Landau-Zener analysis, and $F_{\text{th}}$ is to be regarded as the threshold
field for the Landau-Zener tunnelling. In the presence of dissipation, $F_{\text{th}} = \Delta^2/v$ is no longer valid and 
we need to directly look at Eq.\,(\ref{eq:oka2}).

Note that the imaginary component of the vector potential, $\chi,$ vanishes in 
equilibrium conditions and also in the absence of dissipation. We assume $\chi$ to be a well-behaved 
function of the driving field $F$ near the field-driven transition at $F_{c}$\,:\,$\chi(F_{c})=\chi_{c}.$
It is evident from Eq.\,\eqref{eq:oka2} that for the calculation of
the Landau-Zener tunneling factor $\gamma$, it suffices to obtain
the energy gap for a simpler auxiliary problem with a purely imaginary
vector potential since the expression for $F_{\text{th}}$ is entirely determined by an integral in the $\text{Im}\Psi$ direction. 
The resulting model is non-Hermitian and invariant under
simultaneous parity (${\cal P}$) and time reversal (${\cal T}$)
operations~\cite{tripathi}. For small values of the drive, the eigenvalues of the
${\cal PT}$-symmetric models are real and the corresponding eigenfunctions
are invariant under ${\cal PT}$ transformation. For driving fields
exceeding the critical value, the spectral gap closes, the eigenvalues
acquire finite imaginary parts, and the corresponding eigenfunctions
break ${\cal PT}$ symmetry. More details on the connection between the dynamic Mott transition in dissipative systems and the loss of the ${\cal PT}$ symmetry are given in\,\cite{tripathi}.

\subsection{Dynamic vortex Mott transition near $f_{c}=1$ and ${\cal PT}$ symmetry
	breaking }

We consider now the dynamic vortex Mott transition near the integer
filling $f_{c}=1$, where the vortex Mott transition is described
by an analysis of the nonrelativistic Landau-Ginzburg-Wilson effective
action in Euclidean time, 
\begin{align}
	S & =\int d^{2}x\, d\tau\left[\Psi^{\dagger}\frac{\partial}{\partial\tau}\Psi+D|\nabla\Psi|^{2}+m^{2}|\Psi|^{2}+u|\Psi|^{4}\right].\label{eq:LG-model}
\end{align}
Here $\Psi$ is the vortex field, $D$ the vortex stiffness, $m$
and $u$ are respectively the mass and interaction parameters that
govern the mean-field transition. In mean-field theory, the ``superfluid''
phase of the vortices corresponds to $m^{2}<0.$ In the presence of
the finite electric current, the magnus force on the vortices can be
modeled by incorporating an external vector potential $A_{x}=It,\, A_{y}=0.$
Approaching from the ``normal'' or Mott-insulating side of the vortex
superfluid-Mott insulator transition, we consider the motion of a
vortex in an isolated cell consisting of an Ohmic region bounded by
a large superconducting region, which enables us to impose the simple
boundary condition $\Psi=0$ outside the Ohmic region. If motion in
the Ohmic environment is overdamped, we may assume the time evolution
is entirely governed by Brownian processes and neglect Berry phase
effects (first term in Eq.\,\eqref{eq:LG-model}). We thus get an equation
of motion (in real time) as 
\begin{align}
	\frac{\partial\Psi}{\partial t} & +\rho\frac{\delta H}{\delta\Psi^{*}}=0,\label{eq:eq-motion}
\end{align}
where $H=\int d^{2}x\,\left[D|\nabla\Psi|^{2}+m^{2}|\Psi|^{2}+u|\Psi|^{4}\right]$
is the Hamiltonian corresponding to Eq.\,\eqref{eq:LG-model}, and $\rho$
represents viscous damping of the vortex motion and is phenomenologically
proportional to the (charge) resistivity. Performing gauge transformation
to turn the vector potential into the scalar one, we can recast
Eq.\,\eqref{eq:eq-motion} into the form
\begin{align}
	\frac{\partial\Psi}{\partial t}-i(I/\rho)x\Psi & =D\nabla^{2}\Psi-m^{2}\Psi-2u|\Psi|^{2}\Psi\,.\label{eq:sch-eqn}
\end{align}
For simplicity we ignore the nonlinear term  and consider solutions
of the form $\Psi(x,y,t)=e^{ik_{y}y-\lambda t}u(x)$ with the boundary
conditions $\Psi=0$ in the square superconducting region surrounding
an Au pad which leads us to consider the eigenvalue equation
\begin{align}
	Du_{xx}+i(I/\rho)xu & =-(\lambda-m^{2}-k_{y}^{2})u.\label{eq:linearized}
\end{align}
We can associate a ``Hamiltonian'' ${\cal H=}-Du_{xx}-i(I/\rho)xu$
with Eq.\,(\ref{eq:linearized}) which is evidently ${\cal PT}-$symmetric.
It is easy to see that tuning the current takes us through a ${\cal PT}$ symmetry
breaking phase transition. When $I\rightarrow0,$ the eigenvalues
$\lambda$ are evidently real, and as $I\rightarrow\infty,$ the eigenvalues
$\lambda\sim\pm iIa/\rho.$ The latter limit corresponds to merging
of discrete energy levels since the real part of the eigenvalues has
vanished. We rewrite Eq.\,(\ref{eq:linearized}) in terms of dimensionless
variables $\xi=x/a$ and $E=(\lambda-m^{2}-k_{y}^{2})/E_{T},$ where
$E_{T}=D/a^{2}$ is the Thouless energy:
\begin{align}
	u_{\xi\xi}+i(Ia/E_{T}\rho)u & =-Eu.\label{eq:linearized2}
\end{align}
The critical current $I_{0}$ at which the eigenvalues of equations
of the above form merge has been calculated earlier in the literature~\cite{rubinstein,vinokur}.
Near this bifurcation point, the eigenvalues merge in the following
manner:
\begin{align}
	E_{1}-E_{0} & \approx E_{T}\sqrt{\eta\left(1-\frac{I^{2}}{I_{0}^{2}}\right)}\sim E_{T}\sqrt{\frac{I_{0}-I}{I_{0}}},\label{eq:hopf}
\end{align}
where $\eta\approx(\pi^{2}/\sqrt{2})(I_{0}a/E_{T}\rho)$.  Equation\,(\ref{eq:hopf})
leads to the following scaling of the Landau-Zener tunneling factor:
\begin{align}
	\gamma & \sim(I_{0}-I)^{3/2}.\label{eq:LZ-vortex}
\end{align}

It remains to relate the critical scaling of the Landau-Zener factor
$\gamma$ with the exponent $\epsilon_I$ in the universal scaling
function ${\cal F}.$ In Eq.\,\eqref{eq:oka2}, from the form of the
Landau-Zener factor, $\gamma=I_{\text{th}}/I,$ we identify $I_{\text{th}}$
as a barrier to the generation of free particle-hole pairs. Near the
dynamic Mott transition, we propose that one should relate the potential
energy loss associated with a nearest neighbour hop, $I_{\text{th}}a\sim(I_{0}-I)^{3/2},$
with the Coulomb repulsion $|h|\sim|U_{c}-U|$ associated with the
local correlation. We thus expect the scaling function to be homogeneous
in $|I_{0}-I|/|h|^{2/3}$ near $f_{c}=1,$ and thus $\epsilon_I=2/3.$

\subsubsection{Relation of $\epsilon_I$ with critical exponents $z$ and $\nu$}
From dimensional analysis,
the threshold field for the dynamic Mott transition scales as $I_{\text{th}}\sim|\delta|^{\nu(z+1)}\equiv|I_{0}-I|^{\nu(z+1)},$
where $\nu$ and $z$ refer to the quantum critical theory. Near the
Mott transition, we compare as usual the potential energy change $I_{\text{th}}a$
associated with a nearest-neighbour hop with the local correlation
$|h|$ and arrive at the scaling $|I_{0}-I|/|h|^{1/\nu(z+1)}.$ For
the non-relativistic mean-field case, we use $\nu=1/2$ and $z=2$
and confirm $\epsilon_I=1/\nu(z+1)=2/3.$ 

\subsection{${\cal PT}-$symmetry mechanism of the electric field-driven Mott transition
	in a dissipative fermionic Hubbard chain}

In the discussion above for driven vortex systems, we argued that
a ``non-Hermitian'' imaginary electric field term appears naturally
in the presence of dissipation. The driving current in this case appears
as an electric field acting on the vortex ``charges''. The current-driven
vortex Mott transition is associated with ${\cal PT}$ symmetry breaking.
The idea of ${\cal PT}$ symmetry breaking is also relevant for dynamic
Mott transitions in dissipative fermionic systems. We consider now
a one-dimensional fermionic Hubbard model in the presence of a background
current. As explained in the section about the dielectric-breakdown, in order to
obtain the finite-field renormalization of the spectral gap, one should
consider an auxiliary problem with the purely imaginary vector potential.
This is equivalent to perturbing the equilibrium Hamiltonian $H$
with the current operator $J$ through the Lagrange multiplier\,\cite{antal,cardy},
\begin{align}
	H' & =H-i\lambda J,\label{eq:Lagrange}
\end{align}
where $\lambda$ is real. The model $H',$ while non-Hermitian, has
${\cal PT}-$symmetry if $H$ also has this symmetry, implying a real
spectrum in some parameter range where the eigenstates do not breat
${\cal PT}-$symmetry. In models with the charge conservation,
for example the Hubbard model, the current operator also commutes
with $H$ and one can simultaneously diagonalize $H$ and $J$. It
is easy to see that the nonequilibrium transition is brought about by
tuning $\lambda$. For small $\lambda$ the presence of the spectral
gap means that the expectation value of $J$ (in the model $H^{\prime}$)
vanishes. On the other hand, for very large values of $\lambda$,
the eigenfunctions of $H^{\prime}$ are essentially those of $J$, and a gapless
phase with carrying the finite steady current $I$ becomes possible. The phase
transition from the zero current carrying to the finite current carrying
state thus takes place at the critical value $\lambda=\lambda_{c}$. 

Let us consider the following model for a half-filled fermionic Hubbard chain
subjected to an imaginary gauge field $\chi$: 
\begin{equation}
H^{\prime}=-t\sum_{\langle ij\rangle,\sigma}[e^{\chi}c_{i\sigma}^{\dagger}c_{j\sigma}+e^{-\chi}c_{j\sigma}^{\dagger}c_{i\sigma}]+U\sum_{i}n_{i\uparrow}n_{i\downarrow}.\label{Hubbard-H}
\end{equation}
Rewriting the above Hamiltonian as
\begin{align}
	H^{\prime}= & -t(\cosh\chi)\sum_{\langle ij\rangle,\sigma}[c_{i\sigma}^{\dagger}c_{j\sigma}+c_{j\sigma}^{\dagger}c_{i\sigma}]+U\sum_{i}n_{i\uparrow}n_{i\downarrow}\nonumber \\
	& \qquad-i\sinh(\chi)J,\label{eq:Hubbard-H2}
\end{align}
we can identify $\tanh(\chi)$ with the Lagrange multiplier $\lambda$
in Eq.\,(\ref{eq:Lagrange}) which describes a dissipative model with
the current constraint. To solve Eq.\,(\ref{Hubbard-H}), we utilize
the coupled Bethe ansatz solutions presented in Ref.\,\cite{fukui}
for the charge and spin distribution functions $\rho(k)$ and $\sigma(\lambda)$:
\begin{align}
	\rho(k) & =\frac{1}{2\pi}-\frac{\cos k}{2\pi}\int_{-\infty}^{\infty}d\lambda\,\theta'(\sin k-\lambda)\sigma(\lambda),\label{rho}\\
	\sigma(\lambda) & =-\frac{1}{2\pi}\int_{{\cal C}}dk\,\theta'(\sin k-\lambda)\rho(k)+\nonumber \\
	& \qquad\qquad\frac{1}{4\pi}\int_{-\infty}^{\infty}d\lambda'\,\theta'\left(\frac{\lambda-\lambda'}{2}\right)\sigma(\lambda'),\label{lambda}\\
	\chi(b) & =b-i\int_{-\infty}^{\infty}d\lambda\,\theta(\lambda+i\sinh b)\sigma(\lambda).\label{chi}
\end{align}
Here $\theta(x)=-2\arctan(x/u),$ with $u=U/4t,$ and for $b<b_{\text{cr}}=\mbox{arcsinh}(u),$
the contour ${\cal C}$ is chosen as a pathway in the complex $k$--plane
consisting of the three line segments\,\cite{fukui}, $-\pi+ib\rightarrow-\pi\rightarrow\pi\rightarrow\pi+ib$.
At half-filling, the charge and spin distributions satisfy the constraints
$\int_{{\cal C}}dk\,\rho(k)=1$ and $\int d\lambda\,\sigma(\lambda)=1/2$.
To solve the coupled integral equations for $\rho$ and $\lambda$,
we take the Fourier transform of Eqs.\,(\ref{rho}),\,(\ref{lambda}) and
obtain $\sigma(\lambda)=\int d\omega/2\pi[J_{0}(\omega)/2\cosh(\omega u)]e^{i\lambda\omega}$.
The solution for $\sigma(\lambda)$ is then used in Eq.\,(\ref{chi})
to find the relation between the imaginary gauge field $\chi$ and
the imaginary part of the charge rapidity, $b$. We are particularly
interested in the solution for $b$ near the threshold value $b_{\text{cr}}.$
Let $\chi_{\text{cr}}$ be the imaginary gauge field corresponding
to $b_{\text{cr}}$ in Eq.\,(\ref{chi}). Using our solution for $\sigma(\lambda),$
and taking the derivative with respect to $b$ in Eq.\,(\ref{chi}) we have 
\begin{align}
	\frac{d\chi}{db} & =1-\pi\cosh(b)\int_{-\infty}^{\infty}\frac{d\omega}{2\pi}\frac{J_{0}(\omega)}{\cosh(\omega u)}e^{-\omega\sinh(b)-|\omega|u}\,.\label{chi-soln}
\end{align}
At $b_{\text{cr}},$ it is easy to see that $d\chi/db=0.$ Expanding
the solution around $b=b_{\text{cr}},$ we have $d\chi/db=2C(b-b_{\text{cr}}),$
($C$ is a constant) which gives upon integrating, 
\begin{align}
	\chi_{\text{cr}}-\chi=C(b-b_{\text{cr}})^{2}.\label{chi-b-reln}
\end{align}

The Mott gap is given by\cite{lieb} $\Delta=U-2\mu_{,}$ where 
\begin{equation}
\mu=E(N\downarrow,N\uparrow)-E(N\downarrow-1,N\uparrow)\,.
\end{equation}
At finite $b,$ the Mott gap can be expressed as\,\cite{fukui} 
\begin{align}
	\Delta(b) & =4t\left[u-\cosh(b)+\int_{-\infty}^{\infty}\frac{d\omega}{2\pi}\frac{J_{1}(\omega)e^{\omega\sinh(b)}}{\omega(1+2^{2u|\omega|})}\right].\label{delta}
\end{align}

It is easily checked that $\Delta(b_{\text{cr}})=0,$ and near the
threshold, $\Delta(b)=C'(b_{\text{cr}}-b),$ where $C'$ is a constant.
Combining this with the relation between $\chi$ and $b$ we obtained
in Eq.\,(\ref{chi-b-reln}), we have finally 
\begin{align}
	\Delta(\chi)=A\sqrt{\chi_{\text{cr}}-\chi}.\label{chi-sqroot}
\end{align}
The above $\chi$-dependent Mott gap can be further re-expressed in
terms of the electric field by using the relation $I=\sigma F=\text{tr}(\rho J).$
For well-behaved $I(\chi),$ the same square root singularity $\Delta(F)\sim\sqrt{F_{c}-F}$
is expected for the field dependence of the Mott gap. This leads to
the exponent $\gamma$ vanishing of the threshold field $F_{\text{th}}$ as $\gamma\sim (F_{c}-F)^{3/2}$
near the field-induced transition. For stronger driving fields such
that $\chi>\chi_{\text{cr}},$ the spectrum of the model becomes complex.


\begin{thebibliography}{10}
\bibitem{PM:1937}
N.\,F. Mott and R. Peierls, 
\textit{Discussion of the paper by de Boer and Verwey}, 
Proc. Phys. Soc. \textbf{49} (4S), 72 (1937).

\bibitem{Mott:1949}
N.\,F. Mott, 
\textit{The Basis of the Electron Theory of Metals with Special Reference to the Transition Metals}, Proc. Phys. Soc. A \textbf{62}, 416 (1949).

\bibitem{Mott:1990}
N.\,F. Mott,  
\textit{Metal-Insulator Transitions}. 
(Taylor and Francis, London, 1990).

\bibitem{Sachdev:book}
S. Sachdev, 
\textit{Quantum Phase Transitions}, 2nd ed. 
(Cambridge University Press, 2011).

\bibitem{Polyakov:1987}
A.\,M. Polyakov,
\textit{Gauge fields and strings}, 
(Hardwood Academic Publishers, Chur, 1987) 

\bibitem{Nelson}
D.\,R. Nelson and V.\,M. Vinokur, 
\textit{Boson localization and correlated pinning of superconducting vortex arrays},
Phys. Rev. B \textbf{48}, 13060 (1993).

\bibitem{Zeldov:2009}
S. Goldberg, Y. Segev, Y. Myasoedov, I. Gutman, N. Avraham, M. Rappaport, E. Zeldov, T. Tamegai, C.\,W. Hicks,
and K. A. Moler,
\textit{Mott insulator phases and first-order melting in Bi$_2$Sr$_2$CaCu$_2$O$_{8+\delta}$ crystals with periodic surface holes},
Phys. Rev. B \textbf{79}, 064523 (2009).

\bibitem{Moshchalkov:1998}
V.\,V. Moshchalkov, M. Baert, V.\,V. Metlushko, E. Rosseel, M.\,J. van Bael,K. Temst, Y. Bruynseraede, R. Jonckheere,  \textit{Pinning by an antidot lattice: The problem of the optimum antidot size},
Phys. Rev. B \textbf{57}, 3615 (1998).

\bibitem{Louk:2017}
L. Rademaker, V.\,M. Vinokur, and A. Galda,
\textit{Universality and critical behavior of the dynamical Mott transition in a system with long-range
interactions},
Sci. Rep. \textbf{7}, 44044 (2017).

\bibitem{Imada:1998}
M. Imada, A. Fujimori, and Y. Tokura, 
\textit{Metal-insulator transitions}, 
Rev. Mod. Phys. \textbf{70}, 1039-1263 (1998).

\bibitem{Quantum:2012}
V. Dobrosavljevic, N. Trivedi, and J.\,M. Valles,\,Jr., Ed.
\textit{Conductor-Insulator Quantum Phase Transitions},
 (Oxford University Press, Oxford, UK, 2012).

\bibitem{Balents:2005}
L. Balents, L. Bartosch, A. Burkov, S. Sachdev, and K. Sengupta, \textit{Competing Orders and Non-Landau-Ginzburg-Wilson Criticality
in (Bose) Mott Transitions},
Prog. Theor. Phys. Suppl. \textbf{160}, 314 (2005).

\bibitem{Lee:2006}
P.\,A. Lee, N. Nagaosa, and X.-G. Wen, \textit{Doping a Mott insulator: physics of hightemperature
superconductivity}, Rev. Mod. Phys. \textbf{78}, 17 (2006).

\bibitem{Science2015}
N. Poccia, T.\,I. Baturina, F. Coneri, C.\,G. Molenaar,
X.\,R. Wang, G. Bianconi, A. Brinkman, H. Hilgenkamp,
A.\,A. Golubov, V.\,M. Vinokur,
\textit{Critical behavior at the dynamic vortex insulator to metal transition},
Science \textbf{349}, 1202 (2015).

\bibitem{Nelson:1998}
R.\,A. Lehrer and D.\,R. Nelson,
\textit{Vortex pinning and the non-Hermitian Mott transition},
Phys. Rev. B \textbf{58}, 12385 (1998).

\bibitem{Millis:2006}
A. Mitra, S. Takei, Y.-B. Kim, and A.\,J. Millis,
\textit{Nonequilibrium Quantum Criticality in Open Electronic Systems},
Phys. Rev. Lett. \textbf{97}, 236808 (2006).

\bibitem{Chtchelk:2009}
N.\,M. Chtchelkatchev and V.\,M. Vinokur,
\textit{Nonequilibrium mesoscopic superconductors in a fluctuational regime}, 
Europhys. Lett. \textbf{88}, 407 (2009).

\bibitem{Hubbard:1963}
J. Hubbard,
\textit{Electron correlations in narrow energy bands}, 
Proc. R. Soc. Lond. A \textbf{276}, 238 (1963).

\bibitem{Kotliar:2000}
G. Kotliar, E. Lange, and M.\,J. Rozenberg,
\textit{Landau Theory of the Finite Temperature Mott Transition},
Phys. Rev. Lett. \textbf{84}, 5180 (2000).

\bibitem{Rosenberg}
M.\,J. Rozenberg, R. Chitra, and G. Kotliar,
\textit{Finite temperature Mott transition in the Hubbard model in infinite dimensions}, 
Phys. Rev. Lett. \textbf{83}, 3498 (1999).

%\bibitem{SM} See fabrication details, scaling treatment of the experimental data at additional temperature, and the general scaling analysis of the dynamic Mott transition in Supplementary Materials (SM).
\bibitem{Hebard:1990}
A. Hebard and M.\,A. Paalanen,
\textit{Magnetic-field-tuned superconductor-insulator transition in two-dimensional films},
Phys. Rev. Lett. \textbf{65}, 927 (1990).

\bibitem{Limelette:2003}
P. Limelette, A. Georges, D. J\'{e}rome, P. Wzietek,
P. Metcalf, and J. M. Honig,
\textit{Universality and Critical Behavior at the Mott Transition}, 
Science \textbf{302}, 89 (2003).

\bibitem{Lobb:1983}
 C.\,J. Lobb, D.\,W. Abraham, and M. Tinkham,
\textit{Theoretical interpretation of resistive transition data from arrays of superconducting weak links},
Phys. Rev. B \textbf{27}, 150 (1983).

\bibitem{Mota}
F.\,B. M\"{u}ller-Allinger and A.\,C. Mota,
\textit{Reentrance of the induced diamagnetism in gold-niobium proximity cylinders}, 
Phys. Rev. B \textbf{62}, R6120 (2000).

\bibitem{Vinokur:1998}
V.\,M. Vinokur, B. Khaykovich, E. Zeldov, M. Konczykowski, R.\,A. Doyle,
P.\,H. Kes,
\textit{Lindemann criterion and vortex-matter phase transitions in high-temperature superconductors},
Physica C \textbf{295}, 209 (1998).

\bibitem{Blatter:1994}
G. Blatter, M.\,V. Feigel'man, V.\,B. Geshkenbein, A.\,I. Larkin, and V.\,M. Vinokur,
\textit{Vortices in high-temperature superconductors},
Rev. Mod. Phys. \textbf{66}, 1125 (1994).

\bibitem{Oka:2003}
T. Oka, R. Arita, and H. Aoki,
\textit{Breakdown of a Mott insulator: a nonadiabatic tunneling mechanism}, Phys. Rev. Lett. \textbf{91}, 066406 (2003).

\bibitem{Oka:2010}
M. Eckstein, T. Oka, and P. Werner,
\textit{Dielectric breakdown of Mott insulators in dynamical mean-field theory}, 
Phys. Rev. Lett. \textbf{105}, 146404 (2010).

\bibitem{tripathi:2016}
V. Tripathi, A. Galda, H. Barman, and V.\,M. Vinokur, \textit{Parity-time symmetry-breaking mechanism of dynamic Mott transitions in dissipative systems},
Phys. Rev. B \textbf{94}, 041104\,(R) (2016).

\bibitem{Dykhne:1962}
A.\,M. Dykhne, \textit{Adiabatic perturbation of discrete spectrum states}
JETP \textbf{14}, 941 (1962).

\bibitem{Bender}
C.\,M. Bender and S. Boettcher, \textit{Real Spectra in Non-Hermitian Hamiltonians Having ${\cal PT}$ Symmetry}, 
Phys. Rev. Lett. \textbf{80}, 5243 (1998).

\bibitem{Chtchelk:2012}
N.\,M. Chtchelkatchev, A.\,A. Golubov, T.\,I. Baturina, and V.\,M. Vinokur,
\textit{Stimulation of the Fluctuation Superconductivity by $\cal{PT}$ Symmetry}, 
Phys. Rev. Lett. \textbf{109}, 150405 (2012).


\end{thebibliography}

\begin{thebibliography}{10}
	\bibitem{ceperley}W. R. Magro and D. M. Ceperley, Phys. Rev. Lett.
	\textbf{73}, 826 (1994).
	
	\bibitem{hebert}F. H\`ebert, G. G. Batrouni, R. T. Scalettar, G.
	Schmid, M. Troyer, and A. Dorneich, Phys. Rev. B \textbf{65}, 014513
	(2001).
	
	\bibitem{teitel}See for example S. Teitel and C. Jayaprakash, Phys.
	Rev. B \textbf{27}, 598 (1983) for fully-frustrated 2D Josephson arrays
	at finite temperature.
	
	\bibitem{kotliar}G. Kotliar, E. Lange, and M. J. Rozenberg, Phys.
	Rev. Lett. \textbf{84}, 5180 (2002).
	
	\bibitem{oka}Takashi Oka and Hideo Aoki, Phys. Rev. B \textbf{81},
	033103 (2010).
	
	\bibitem{hatano}N. Hatano and D. R. Nelson, Phys. Rev. Lett. \textbf{77},
	570 (1996); N. Hatano and D. R. Nelson, Phys. Rev. B \textbf{56},
	8651 (1997).
	
	\bibitem{albertini}Giuseppe Albertini, Silvio Renato Dahmen and Birgit
	Wehefritz, J. Phys. A: Math. Gen. \textbf{29}, L369 (1996).
	
	\bibitem{rubinstein}J. Rubinstein, P. Sternberg, and Q. Ma, Phys.
	Rev. Lett. \textbf{99}, 167003 (2007).
	
	\bibitem{vinokur}N. M. Chtchelkatchev, A. A. Golubov, T. I. Baturina,
	and V. M. Vinokur, Phys. Rev. Lett. \textbf{109}, 150405 (2012).
	
	\bibitem{tripathi}
	V.\,Tripathi, A.\,Galda, H.\,Barman,
	V.\,M.\,Vinokur, Phys. Rev. B \textbf{94}, 041104(R) (2016).
	
	\bibitem{brezin}E. Br\'ezin and D. J. Wallace, Phys. Rev. B \textbf{7},
	1967 (1973).
	
	\bibitem{pelissetto}A. Pelissetto and E. Vicari, Phys. Rep. \textbf{368},
	549 (2002).
	
	\bibitem{klimenko}K. G. Klimenko, Z. Phys. C \textbf{54}, 323-329
	(1992).
	
	\bibitem{antal}T. Antal, Z. R\'{a}cz, and L. Sasv\'{a}ri, Phys.
	Rev. Lett. \textbf{78}, 167 (1997).
	
	\bibitem{cardy}John Cardy and Peter Suranyi, Nucl. Phys. B \textbf{565},
	487 (2000).
	
	\bibitem{fukui}T. Fukui and N. Kawakami, Phys. Rev. B \textbf{58},
	16051 (1998).
	
	\bibitem{lieb}E. H. Lieb and F. Y. Wu, Phys. Rev. Lett. \textbf{20},
	1445 (1968). \end{thebibliography}
\end{document}